\documentclass[12pt,times]{article}
\title{Active- and transfer-learning applied to microscale-macroscale coupling to simulate viscoelastic flows}

\author{Lifei Zhao$^{1,2}$, Zhen Li$^{3}$\footnote{Corresponding Authors. Email: \href{mailto:zli7@clemson.edu}{zli7@clemson.edu}~(Zhen Li)}, Zhicheng Wang$^2$, Bruce Caswell$^4$,\\ Jie Ouyang$^{1}$ and George Em Karniadakis$^2$\footnote{Email: \href{mailto:george_karniadakis@brown.edu}{george\_karniadakis@brown.edu}~(George Em Karniadakis)} \vspace{-0.1cm}\\
\small{$^1$~Department of Applied Mathematics, Northwestern Polytechnical University, Xi'An 710129, China}\vspace{-0.1cm}\\
\small{$^2$~Division of Applied Mathematics, Brown University, Providence, RI 02912, USA}\vspace{-0.1cm}\\
\small{$^3$~Department of Mechanical Engineering, Clemson University, Clemson, SC 29634, USA}\vspace{-0.1cm}\\
\small{$^4$~School of Engineering, Brown University, Providence, RI 02912, USA} \vspace{-0.75cm} }

\date{}

\RequirePackage{lineno}
\usepackage[colorlinks=true]{hyperref}
\usepackage{graphicx}
\usepackage{multirow,array,color,mathrsfs,bm,subcaption}
\usepackage{epstopdf}
\usepackage[noadjust,compress]{cite}

\usepackage{caption}
\DeclareGraphicsExtensions{.eps,.mps,.pdf,.jpg,.png}
\graphicspath{{figures/}{../figures/}}

\usepackage{amsfonts,amsmath,amssymb}

\usepackage{dsfont}

\usepackage[top=1.5cm,bottom=2.25cm,left=1.75cm,right=1.75cm]{geometry}

\usepackage[figuresright]{rotating}
\usepackage{extarrows}

\begin{document}

\maketitle

\begin{abstract}
Active- and transfer-learning are applied to microscale dynamics of polymer flows for the multiscale discovery of effective constitutive approximations required in viscoelastic flow simulation. The result is macroscopic rheology directly connected to a microstructural model. Micro and macroscale simulations are adaptively coupled by means of Gaussian process regression (GPR) to run the expensive microscale computations only as necessary. This multiscale method is demonstrated with flows of a polymer solution as a model system. At the microscale level dissipative particle dynamics (DPD) is employed to model the fluid as a suspension of bead-spring micro-structures subjected to steady shear flow. The results yield the non-Newtonian viscosity and the first normal stress difference at strain rates as training data used in a GPR model. DPD parameters are calibrated with respect to experimental data for a real polymer solution. Compliance with these data requires adjustment of the DPD model's cutoff radius, which then becomes a function of the second invariant of the strain rate tensor.
The FENE-P model is chosen for the macroscale description using the spectral element method (SEM) to simulate channel flow and flow past a circular cylinder. The DPD results at the lowest possible shear strain rate yield an estimate of the zero-shear rate viscosity, which allows the initiation of the macroscale flow by SEM as a Newtonian fluid. The resulting strain-rate field is surveyed to determine additional shear strain rate sampling points for the DPD system. This new information allows an initial fitting of parameters of the constitutive equation followed by new SEM simulations at the macroscale. Guided by active-learning GPR to select new sampling points, this process continues until convergence is achieved.

The effectiveness of this new simulation paradigm for viscoelastic flows is tested with different macroscale operating conditions. The effective closure learned in the channel simulation is then transferred directly to the flow past a circular cylinder at low Reynolds number, where the results show that only two additional DPD simulations are required to achieve a satisfactory constitutive model. With an increase of the Reynolds number, the active-learning scheme automatically detects the inaccuracy of the learned constitutive model, and initiates additional DPD simulations for the extra data needed to once again close the microscale-macroscale coupled system.
This new paradigm of active- and transfer-learning for multiscale modeling is readily applicable to other microscale-macroscale coupled simulations of complex fluids and other materials. Furthermore, the coupling between microscale and macroscale solvers can be seamlessly implemented with our open source multiscale universal interface (MUI) library.
\end{abstract}

\section{Introduction}
Simulation of viscoelastic flows at the continuum or macroscale level requires a constitutive equation for the stress tensor whose parameters have traditionally been determined from appropriate physical rheological tests. This investigation explores an alternative to such expensive tests, namely the use of microscale numerical models, which achieve lowest cost with on-the-fly strategies that minimize the number of expensive micro-simulations required to arrive at a satisfactory constitutive equation.
In steady shear flows, a Newtonian fluid exhibits normal stresses only as an isotropic pressure, whereas in polymeric liquids the normal stresses are generally anisotropic, an indication of viscoelasticity~\cite{2010Sunthar}. The need to simulate flows of polymeric materials in different geometries under varying operating flow conditions motivates the development of robust constitutive equations, which relate the stress tensor to appropriate kinematic measures of deformation. Over the years many such equations have been proposed and tested experimentally, with their parameters determined by subjecting samples to shear and stretching tests~\cite{2015Larson}. In the present work, the `samples' become the microscopic dissipative particle dynamics (DPD) models of polymer solutions, and the tests are limited to steady shear flow. Unsteady shearing test flows remain uneconomical with DPD, and likewise stretching flows are limited to the study of single chains in suspension~\cite{2016Yazdani}. In view of the limitation on rheological testing, the  model fluid in this work is a dilute polymer solution whose stress tensor is determined by the FENE-P constitutive equation. It was originally derived by Peterlin~\cite{1980Bird} as a dilute suspension of elastic dumbbells, and was used by Purnode and Crochet~\cite{1996Purnode,1998Purnode} to represent rheological test data for a dilute aqueous polyacrylamide solution.
In previous work~\cite{2018Zhao} a multiscale method was demonstrated to efficiently simulate flows of the generalized non-Newtonian fluid characterized only by the shear viscosity function calculated from microscale simulations. Here, both the viscosity and the first normal stress difference are obtained from the same simulations, which allows this method to be extended to viscoelastic fluids by deriving the FENE-P parameters from those functions of the shear rate. The goal is to calculate these material functions of macroscopic rheology only over the relevant range of shear rates to minimize the number of expensive microscale simulations while maintaining sufficient accuracy to construct the constitutive equation required for the simulation of the macroscopic flow under the specified operating conditions.

First, a DPD model of the polymer solution is constructed to capture its microscale dynamics in steady shear flows. Here, we assume homogeneous flow conditions, and thus we neglect the flow-induced concentration changes because of stress-induced polymer migration~\cite{2013Germann}. After calibration of DPD parameters by dimensional scaling to bring the microscale simulations into agreement with realistic physical data, the non-Newtonian viscosity and the first normal stress difference can be derived as functions of the shear strain rate as needed.
In this work the experimental rheological values for an aqueous polyacrylamide solution~\cite{1996Purnode,1998Purnode} serve as the calibration data for the model polymer solution.
The result is a DPD model with realistic steady shear rheology ready for the fitting of FENE-P parameters. Next, the FENE-P constitutive equation is solved together with the conservation equations by the spectral element method (SEM). Guided by active- and transfer-learning the solution proceeds by advancing the operational driving force, and by concurrently running the DPD simulations only as necessary to yield the effective model parameters at shear strain rates optimally chosen by the active- and transfer-learning scheme.

\begin{figure}[t!]
\centering
\includegraphics[width=0.85\textwidth]{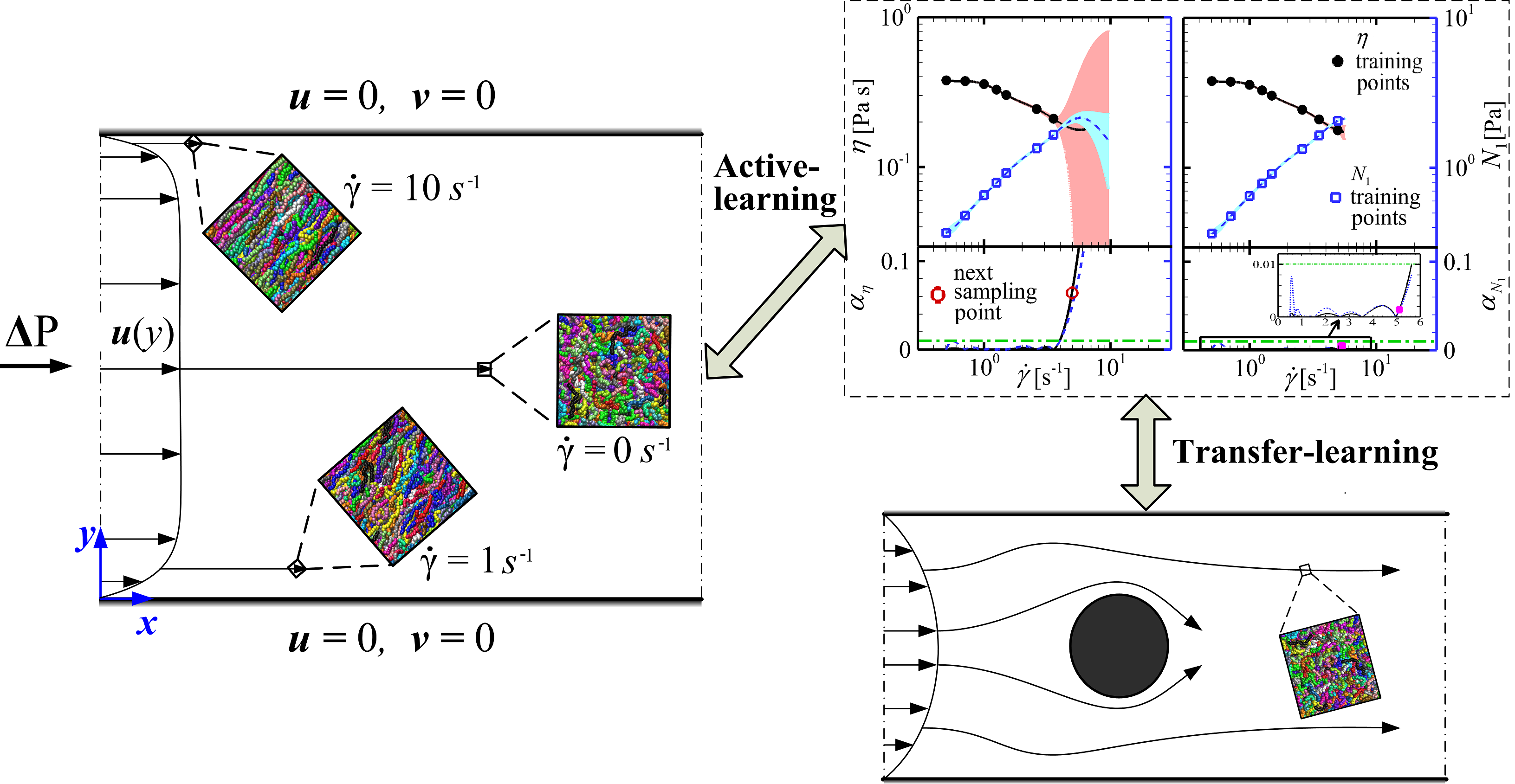}
\caption{Schematic illustration of how active- and transfer-learning combines micro- and macro-scale models to construct effective constitutive models directly from microscale dynamics. The square subdomains within the channel
are the sites for DPD simulations, leading to viscosity and normal stress inferred values
(with error bounds shown by the shaded areas) in the upper right.
Active-learning is implemented by adding selected sampling points (square domains)  while concurrently running microscale simulations to adaptively construct an effective constitutive model, which once learned for one viscoelastic flow can be directly transferred to other flow problems in different domains or to the same flow at different operating conditions (see lower right).}

\label{FIG:figure1}
\end{figure}

Gaussian process regression (GPR) is particularly effective for exploring and exploiting parameter space for a relatively small number of parameters while quantifying uncertainty. The latter is then used to construct proper acquisition functions, whose minima lead to simple active-learning algorithms~\cite{2018Zhao,2019Fan}. A schematic of this process for the current problems is shown in Fig.~\ref{FIG:figure1}. Improved agreement between the simulated viscosity and normal stress functions and the physical values is found to be achieved by allowing certain FENE-P parameters to vary with the shear strain rate, similarly to the White-Metzner model~\cite{1963White}.
After the shear-thinning viscosity and relaxation time are computed from a few selected DPD simulations, the pressure driven channel flow proceeds with the active-learning scheme for microscale-macroscale coupled simulations. How optimal sampling points are adaptively selected on-the-fly while being informed by GPR will be explained in detail below. For flows of the same viscoelastic fluid, the learned constitutive equation for a particular flow can be directly transferred to other flow problems in different domains or to the same problem at different operating conditions. Hence, the GPR model actively learned in a channel flow at one flow rate can be transferred to another operating at a different flow rate, and further to a new geometry with viscoelastic flow past a circular cylinder.

The remainder of this paper is organized as follows: In Section~\ref{sec:2} the details are given for the particle DPD model and the continuum FENE-P model of a polymer solution. In Section~\ref{sec:2_1} this includes a DPD model with the range of interaction cutoff between particles allowed to vary with the local strain-rate invariant, and so to capture the shear-thinning viscosity and the normal stress difference functions consistent with the experimental data for an aqueous polyacrylamide solution. Likewise the FENE-P model is modified with a strain-rate dependent relaxation time necessary to generate bulk rheology consistent with experiment. Then in Section~\ref{sec:2_2} the continuum system for the FENE-P constitutive equation together with the conservation laws for momentum and mass are introduced. The system is discretized with quadrilateral elements and solved by SEM. To stabilize the SEM solver, an extra nonlinear diffusion term, analogous to the entropy-viscosity term used in aerodynamics, is added to the FENE-P equation. In Section~\ref{sec:3}, the effectiveness of using the GPR based active- and transfer-learning scheme is demonstrated with numerical examples of microscale-macroscale coupled simulations of viscoelastic flows. Finally, we conclude with a brief summary and discussion in Section~\ref{sec:4}.

\section{Methods}\label{sec:2}
\subsection{Microscale Method}\label{sec:2_1}
Relations between the stress tensor and kinematic measures of deformation for non-Newtonian polymeric fluids are generally complicated, and reflect the collective dynamical response of polymer micro-structures to macroscopic deformation~\cite{2002Tucker}.
In the present study, the DPD method is employed to explicitly simulate the dynamics of polymer micro-structures at a coarse-grained level. Similar to molecular dynamics (MD) systems, a DPD model consists of many interactive particles governed by the Newton's equation of motion~\cite{1997Groot},
\begin{equation}\label{eq:EOM_DPD}
m_i\frac{{\rm d}^2\mathbf{r}_i}{{\rm d}t^2} = m_i\frac{{\rm d}\mathbf{v}_i}{{\rm d}t} = \mathbf{F}_i = \sum_{j\ne i}\left(\mathbf{F}^C_{ij} + \mathbf{F}^D_{ij} + \mathbf{F}^R_{ij}\right),
\end{equation}
where $m_i$ is the mass of a particle $i$ and $t$ denotes time. The bold symbols $\mathbf{r}_i$ and $\mathbf{v}_i$ represent position and velocity vectors, and $\mathbf{F}_i$ is the total force acting on the particle $i$ due to the presence of neighboring particles. The summation for computing $\mathbf{F}_i$ is carried out over all other particles within a cutoff radius beyond which the pairwise interactions are negligible.
The pairwise force is comprised of a conservative force $\mathbf{F}^C_{ij}$, a dissipative force $\mathbf{F}^D_{ij}$ and a random force $\mathbf{F}^R_{ij}$, which respectively are of the forms: \vspace{-0.2cm}
\begin{equation}\label{eq:F_DPD}
\begin{split}
   & \mathbf{F}^C_{ij} = a_{ij}\omega_C(r_{ij})\mathbf{e}_{ij}, \\
   & \mathbf{F}^D_{ij} = -\gamma_{ij}\omega_D(r_{ij})(\mathbf{v}_{ij}\mathbf{e}_{ij})\mathbf{e}_{ij}, \\
   & \mathbf{F}^R_{ij}\cdot {\rm d}t = \sigma_{ij}\omega_R(r_{ij}){\rm d}\tilde{W}_{ij}\mathbf{e}_{ij},
\end{split}
\end{equation}
where $r_{ij}=|\mathbf{r}_{ij}| = |\mathbf{r}_i-\mathbf{r}_j|$ is the distance between two particles $i$ and $j$, $\mathbf{e}_{ij}=\mathbf{r}_{ij}/r_{ij}$ is the unit vector, and $\mathbf{v}_{ij} = \mathbf{v}_i-\mathbf{v}_j$ is the difference of their velocities; ${\rm d}\tilde{W}_{ij}$ is an independent increment of the Wiener process~\cite{1995Espanol}. The coefficients $a_{ij}$, $\gamma_{ij}$ and $\sigma_{ij}$ determine the strength of each force respectively. To satisfy the fluctuation-dissipation theorem (FDT)~\cite{1995Espanol}, and to maintain the DPD system at a constant temperature~\cite{1997Groot}, the dissipative force and the random force are related by $\sigma_{ij}^2 = 2\gamma_{ij}k_BT$ and $\omega_D(r_{ij})=\omega^2_R(r_{ij})$.
In the classical DPD model, all the forces in Eq.~\eqref{eq:F_DPD} have the same finite interaction range $r_c$ and their amplitudes decay with the distance $r_{ij}$ according to corresponding weight functions $\omega_C(r_{ij}) = 1 - r_{ij}/r_c$, $\omega_D(r_{ij}) = \omega^2_R(r_{ij}) = (1 - r_{ij}/r_c)^{1/2}$ with $r_c$ being the cutoff radius.

The micro-structures of polymer molecules are represented as bead-spring chains so that their dynamics can be explicitly simulated when their suspensions are subject to flows. The chain structure of a polymer is modeled as $N$ beads connected by a finitely extensible nonlinear elastic (FENE) spring~\cite{1990Kremer}. The FENE potential is given by
\begin{equation}\label{eq:FENE_DPD}
U_{B}(r)=\left\{\begin{matrix}
 -\frac{1}{2}k_s r_0^2 \ln \left [ 1 - (r/r_0)^2 \right ] \ ,  &r \leq r_0 \\
 \infty \ ,  &r > r_0 \\
\end{matrix}\right.
\end{equation}
where $k_s$ is the spring constant and $r_0$ the maximum bond extension of the FENE spring. When chosen to be short and stiff enough the nonlinear FENE bond has been shown to minimize phantom bond-crossings~\cite{2007Zhou} during chain entanglements.

The polymer solution is then represented by many bead-spring chains suspended in a solvent of free DPD particles, a widely used model in studies of polymer solutions~\cite{2010Fedosov,2015LiZ,2016Litvinov,2016Lisal}. DPD is a coarse-grained method whose input parameters must be prescribed, and whose macroscopic rheological properties, such as viscosity and diffusivity, are outputs. However, it is worth noting that the DPD model is grounded in atomistic dynamics and can be derived directly from MD systems using the rigorous mathematical framework of the Mori-Zwanzig projection~\cite{2014LiZ}. In our demonstration we calibrate the DPD model of polymer solution with experimental values measured at various shear strain rates by Purnode and Crochet~\cite{1996Purnode} for a 0.5\% aqueous polyacrylamide solution. We start with the classical DPD model with constant cutoff radius for pairwise interactions. Specifically, we consider a DPD system with a number density of $\rho_n = 3.0$ at $k_BT = 1.0$, and set the parameters in Eq.~\eqref{eq:F_DPD} as $a_{ss}=a_{pp} = 25.0,~a_{ps}=12.5,~\gamma_{ij} = 4.5, \sigma_{ij} = 3.0, r_c = 1.0$, where the subscripts `$s$' represents solvent and `$p$' for polymer.
The polymer solution consists of $756$ polymer chains suspended in $88\,200$ DPD solvent particles in a computational box of $60\times70\times10$ (in reduced DPD units). Simple-shear DPD simulations at various shear strain rates are performed to obtain the shear stress and the first normal stress difference computed by the Irving-Kirkwood formula~\cite{2012Yang}.

\begin{figure}[t!]
\centering
\includegraphics[width=0.8\textwidth]{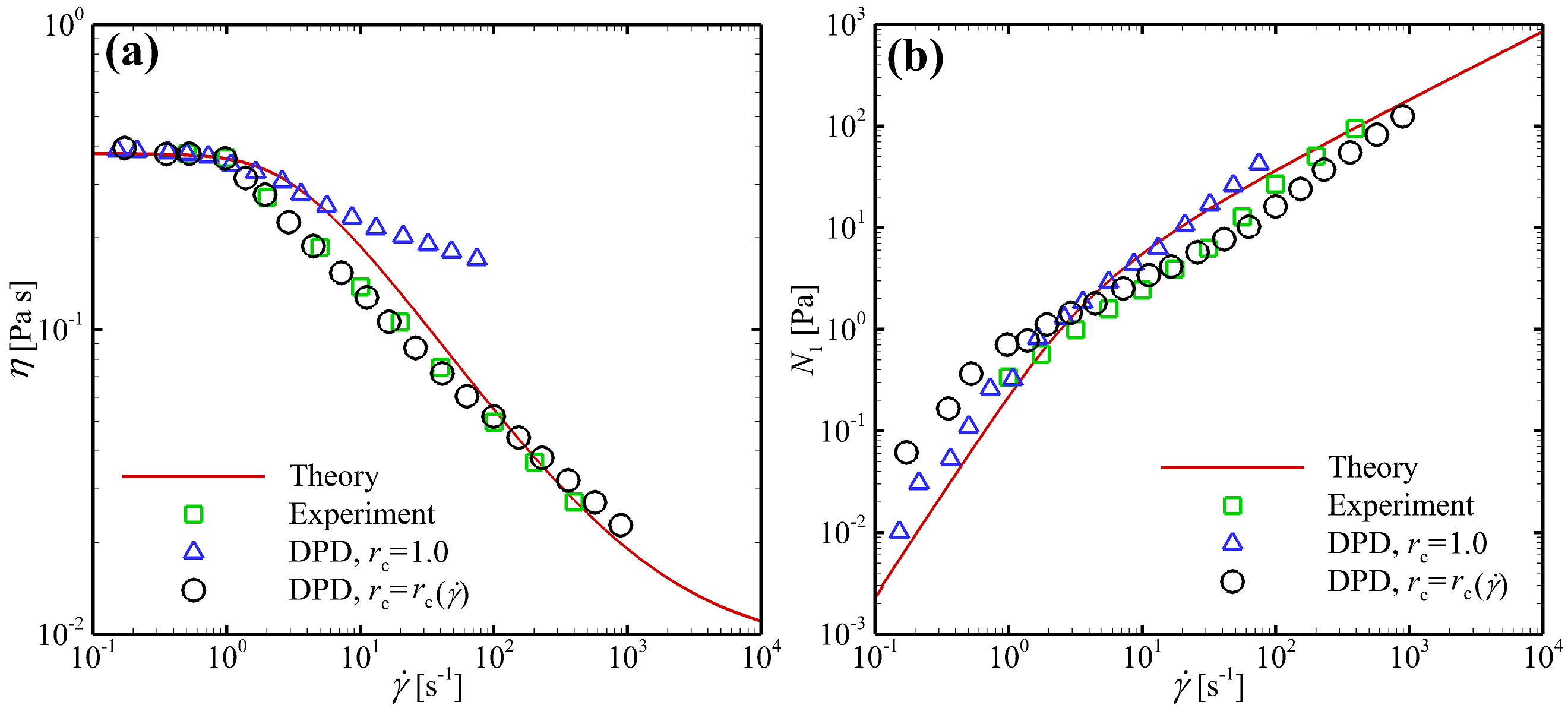}
\caption{Validation of the DPD model with  variable cutoff radius Eq.~\eqref{eq:rc_DPD}  (a) Non-Newtonian viscosity $\eta$ and (b) first normal stress difference $N_1$ as functions of strain-rate $\dot\gamma$ in steady shear flow computed by the modeled DPD polymer solution compared to experimental data for a $0.5\%$ aqueous polyacrylamide solution  ~\cite{1996Purnode} and ``Theory" (red lines) derived from the FENE-P model (Eq. 5). DPD results: for constant cutoff radius, triangles $r_c=1.0$, and for variable cutoff radius (circles) Eq.~\eqref{eq:rc_DPD}. \vspace{-0.2cm}}
\label{FIG:figure2}
\end{figure}

The DPD results are mapped to the corresponding measured physical ones in the experimental data by choosing appropriate characteristic length, mass and time units. For the standard DPD model with constant cutoff radius, this mapping yields $[L] = 1.138\times10^{-3}~m$, $[M]= 4.92\times10^{-7}~kg$ and $[T]=5.43\times10^{-3}~s$ as the characteristic length, mass and time units, respectively.
Figure~\ref{FIG:figure2} presents the rheological properties of the polymer solution obtained by the standard DPD model of bead-spring chains suspended in a solvent of free particles with $r_c=1.0$ compared to experimental data for a 0.5\% aqueous polyacrylamide solution~\cite{1996Purnode}.
It is obvious from Fig.~\ref{FIG:figure2} that the DPD model with constant cutoff radius $r_c = 1.0$ correctly captures the shear-thinning feature of polymer solution, however, the magnitude of viscosity variation in the DPD model is one-order smaller than the experimental data, as shown in Fig.~\ref{FIG:figure2}(a). The apparent reason appears to be that the DPD model is a coarse-grained representation of the polymer molecules which does not model their atomistic structures. Each bead in a DPD chain represents a cluster of many monomers, where the changes of unresolved polymer structures due to shear are not captured well by the DPD model. Polymer coils become stretched out under increasing shear strain rate. The range of shear strain rates for stretch-out is far wider for real polymer coils than that found in simulations with the coarse-grained polymer model. This may explain why the latter fails to capture correct shear-thinning effect. To remedy this, we take the cutoff radius to be a function of the second invariant of the strain rate tensor ${\bm\epsilon}$, in the form of
\begin{equation}
\label{eq:rc_DPD}
r_c(\dot\gamma) = r_{c,0} + \kappa\log\left(\frac{\chi}{\dot\gamma}\right),
\end{equation}
where $\dot{\gamma} = \sqrt{2{\rm{tr}}({\bm\epsilon}^2)}$ represents the second invariant of the strain rate tensor; $r_{c,0}$, $\kappa$ and $\chi$ are undetermined parameters. Eq.~\eqref{eq:rc_DPD} assumes that the interaction range of polymer beads decreases with increasing shear strain rates. Because the transport properties such as viscosity and diffusivity of a DPD system are determined mainly by the dissipative and random forces, we set the $r_c$ in $\mathbf{F}^D_{ij}$ and in $\mathbf{F}^R_{ij}$ as a function of $\dot\gamma$ in Eq.~\eqref{eq:rc_DPD}, while we maintain the $r_c$ in $\mathbf{F}^C_{ij}$ constant as $r_c = r_{c,0}$. In particular, the variable cutoff radius $r_c(\dot\gamma)$ is bounded by the region [1,2] to avoid unreasonable large or small cutoff radii, i.e., $r_c(\dot\gamma) = \max[1,\min(2, r_c(\dot\gamma))]$. We set the parameters in Eq.~\eqref{eq:rc_DPD} as $r_{c,0}=1.0$, $\kappa = 0.265$ and $\chi = 10.0$ in reduced DPD units to generate rheological properties consistent with the experimental data.

For the DPD model with variable cutoff radius, the reduced DPD units are mapped to physical units and compared to experimental data by setting $[L] = 1.44\times10^{-4}~{\rm m}$, $[M]=9.99\times10^{-10}~{\rm kg}$ and $[T]=1.1\times10^{-3}~{\rm s}$ as the characteristic length, mass and time units, respectively.
Then, all other physical units of flow and rheological variables can be mapped to reduced DPD units by dimensional analysis using the three basic units $[T]$, $[L]$ and $[M]$, i.e., the shear strain rate is converted by the strain rate unit $[\dot\gamma]=1/[T]=903.22~{\rm s}^{-1}$. Also, the dynamic viscosity and the shear stress are converted by the factors $[\eta]=[M][L]^{-1}[T]^{-1}=6.304\times10^{-3}~{\rm Pa\,s}$, $[\tau]=[M][L]^{-1}[T]^{-2}=5.694~{\rm Pa}$, and $[N_1]=2[M][L]^{-1}[T]^{-2}=11.39~{\rm Pa}$, respectively.
Figure~\ref{FIG:figure2} shows the dependence of the non-Newtonian viscosity $\eta$ and the first normal stress difference $N_1$ on shear strain rate $\dot\gamma$ in steady shear flow. The DPD results with variable cutoff radius $r_c(\dot\gamma)$ given by Eq.~\eqref{eq:rc_DPD} are now consistent with the experimental data for a 0.5\% aqueous polyacrylamide solution~\cite{1996Purnode}.

In a steady shear flow, analytical expressions for the FENE-P fluid were derived by Purnode and Crochet~\cite{1998Purnode} for the strain rate dependent viscosity $\eta$, and the first normal stress difference $N_1$, as
\begin{equation}\label{eq:shear_FENE}
\begin{split}
\eta&=\eta_s+\eta_p\left(\frac{L^6}{4\lambda^2\dot{\gamma}^2\left(L^2-3\right)^2}\right)^{1/3}\left(\Delta_1^{1/3}+\Delta_2^{1/3}\right),
\\
N_1&=\frac{\eta_pL^4}{\lambda\left(L^2-3\right)}\left(\frac{2L^6}{\lambda^2\dot{\gamma}^2\left(L^2-3\right)^2}\right)^{-1/3}\left(\Delta_1^{1/3}+\Delta_2^{1/3}\right)^2,
\end{split}
\end{equation}
where $\eta_s$ and $\eta_p$ are the solvent and polymer viscosities respectively, $\lambda$ is the relaxation time, and $L$ is the ratio of the extended length of the spring to its equilibrium length.
Hence $L^2$ is a measure of the extensibility of the FENE dumb-bells.
$\Delta_1=1+[1+2L^6/(27\lambda^2\dot{\gamma}^2(L^2-3)^2)]^{1/2}$, and $\Delta_2=2-\Delta_1$. Further details of the FENE-P fluid will be introduced in Section~\ref{sec:2_2}.

Purnode and Crochet~\cite{1998Purnode} fitted their experimental data plotted in Fig.~\ref{FIG:figure2} using a FENE-P parameter set $\eta_s=0.012~{\rm Pa\,s},~\eta_p=0.378~{\rm Pa\,s},~L^2=4,~\lambda=2~s$. Although their non-Newtonian viscosities are well-fitted, the normal stress differences deviate significantly from the FENE-P curve, i.e., the opposite sign of the second derivative of the curves, which suggests that the 0.5\% aqueous polyacrylamide solution is not a perfect FENE-P fluid. Likewise it also shows that the standard bead-spring DPD model does not yield the FENE-P stress even when chain lengths are varied substantially. The improved fits of both properties with the introduction of a variable cutoff radius $r_c(\dot\gamma)$ dependent on a macro level measure of deformation rate indicates that modeling mesoscopic inter-molecular forces is a crucial factor in obtaining realistic macroscopic viscoelastic constitutive equations.

\subsection{Continuum Method}\label{sec:2_2}
The flow of an incompressible fluid is governed by the conservation of mass and momentum~\cite{2013Karniadakis},
\begin{equation}\label{eq:Gov}
\begin{split}
  \nabla\cdot \mathbf{u} &= 0, \\
  \rho\frac{\partial \mathbf{u}}{\partial t} +\rho \mathbf{u}\cdot\nabla\mathbf{u} = &-\nabla p+\nabla\cdot\bm{\tau},
\end{split}
\end{equation}
where $t$, $\rho$ and $p$ denote time, density and pressure, respectively; $\mathbf{u}$ is the velocity vector, and $\bm{\tau}$ the stress tensor. The momentum equation has to be closed by a specified constitutive equation. This work is motivated by the flows of polymer solutions, hence the chosen constitutive equation is FENE-P ~\cite{2005Cruz,2015Jafari} given by,
\begin{equation}\label{eq:fenep_constitutive}
\bm{\tau}=\bm{\tau^s}+\bm{\tau^p},~~~\bm{\tau}^s=\eta_s\cdot 2{\bm \epsilon},~~~\bm{\tau}^p=\frac{\eta_p}{\lambda}\frac{L^2}{L^2-3}\left(\frac{L^2-3}{L^2-tr\left(\bm{C}\right)}\bm{C}-\bm{I}\right),
\end{equation}
where $\bm{\tau}^s$ is the stress generated by the solvent with $\eta_s$ being the solvent viscosity, ${\bm\epsilon}=[\nabla\mathbf{u}+(\nabla\mathbf{u})^T]/2$ is the rate of deformation tensor, $\bm{\tau}^p$ is the stress generated by the polymer with $\eta_p$ being the polymer viscosity, $\lambda$ is the relaxation time, $L$ is the extensibility of the FENE dumbbell, and $\bm{C}$ is the conformation tensor computed from the following equation,
\begin{equation}\label{eq:C}
\overset{\triangledown}{\mathbf{C}}=-\frac{1}{\lambda}\frac{L^2}{L^2-3}\left(g\mathbf{C}-\mathbf{I}\right),~~~g=\frac{L^2-3}{L^2-tr\left(\mathbf{C}\right)},
\end{equation}
in which $\overset{\triangledown}{\mathbf{C}}$ is the Oldroyd upper convected derivative defined by
\begin{equation}
{\overset{\triangledown}{\mathbf{C}}=\frac{\partial \mathbf{C}}{\partial t}+\mathbf{u} \cdot \nabla \mathbf{C}-\mathbf{C} \cdot (\nabla \mathbf{u}) - (\nabla \mathbf{u})^T \cdot \mathbf{C}.}
\end{equation}
In order to improve the performance of the continuum model in reproducing the experimental observations, instead of using a classical FENE-P model with constant relaxation time and polymer viscosity, we propose a modified FENE-P model with polymer viscosity and relaxation-time dependent on the second invariant of strain rate by drawing an analogy to the White-Metzner model~\cite{1992Maders} with variable viscosity and relaxation time, i.e., $\eta_p(\dot\gamma)$ and $\lambda(\dot\gamma)$, which can be obtained from simple shear DPD simulations of polymer solution at various shear strain rates.

Eqs.~\eqref{eq:Gov}-\eqref{eq:C} are non-dimensionalized by the characteristic length $L_0$, velocity $U_0$ and viscosity $\eta_0$. All physical variables are scaled accordingly as: $x^*=x/L_0$, $u^*=u/U_0$, $\eta^*=\eta/\eta_0$, $t^*=tU_0/L_0$, $p^*=p/(\rho U_0^2)$ and $\tau^*=\tau L_0/(\eta_0 U_0)$.
The non-dimensionalized Eq.~\eqref{eq:Gov}-\eqref{eq:C} were solved using the SEM method~\cite{2013Karniadakis} with a high-order time-splitting method with stiffly-stable scheme~\cite{2003Ma} for temporal discretization. Moreover, we employed the entropy viscosity method (EVM)~\cite{2019Wang} that introduces a nonlinear diffusion term to the FENE-P constitutive equation to stabilize the SEM simulation at high Weissenberg numbers (see details in Appendix~\ref{sec:app}).

\subsection{Multiscale Coupling}\label{sec:2_3}
A polymer solution can exhibit non-Newtonian (shear-thinning) behavior, i.e., the shear viscosity and normal stress differences change with shear strain rate or shear strain rate history. The reason is that the polymer molecules suspended in solvent have chain like micro-structures, namely the length along the polymer chain is much larger than the other dimensions of the molecule. More importantly, the micro-structure of polymer is deformable and can be significantly changed by the flow conditions. The dissolved polymer molecules are generally in a random coil state in an equilibrium solution~\cite{2010Fedosov,2018ZLi}. However, under shear flows, the polymer chains trend to be aligned in the direction of shear and generate a reduced flow resistance as the shear strain rate increases. Consistently, the apparent viscosity of a polymer solution will decrease with increasing shear strain rate and exhibits a shear thinning behavior, while the normal stress differences increase as the shear strain rate increases. Using multiscale simulations, we couple the microscale and macroscale descriptions and directly connect the dynamics of polymer chains and the continuum equation for bulk rheology. Hence, a constitutive closure model can be on-the-fly constructed from the microscale dynamics using an active-learning algorithm rather than using an {\em ad hoc} empirical formula for the constitutive relation.

At the microscale level, the polymer solution is modeled by bead-spring chains suspended in a solvent of free DPD particles. Specifically, the DPD system is constructed with $756$ chains in a computational box of $60\times70\times10$ in reduced DPD units. Each chain consists of $50$ DPD beads connected by the FENE spring given by Eq.~\eqref{eq:FENE_DPD} with spring constant $k_s=30$ and $r_0=1.5$. The overall particle number density of the system is $\rho_n=3$ and the temperature is $k_BT = 1.0$. The non-bonded interactions between DPD particles are computed by Eq.~\eqref{eq:F_DPD}. For the conservative force, we use $a_{ss}=a_{pp}=25, a_{ps}=12.5$, and $\omega_C(r)=1-r/r_c$ with a cutoff radius $r_c = 1.0$, where the subscript `$s$' represents solvent and `$p$' for polymer. For the dissipative force, we use $\gamma_{ss}=\gamma_{pp}=\gamma_{ps}=4.5$ and $\omega_D(r)=[1-r/r_c(\dot\gamma)]^{1/2}$ with a variable cutoff radius $r_c(\dot\gamma)$ given by Eq.~\eqref{eq:rc_DPD}. The parameters in the random force are given by $\sigma_{ij}^2 = 2\gamma_{ij}k_BT$ and $\omega^2_R(r)=\omega_D(r)$. We will perform simple shear DPD simulations at a few selected shear strain rates $\dot\gamma$ to obtain the stress tensor~$\bm\tau$ at various shear strain rates, and we will compute the shear tensor is computed by the Irving-Kirkwood formula~\cite{2012Yang}. Let $\tau_{xy}$ be the shear stress and $N_1=\tau_{xx}-\tau_{yy}$ be the first normal stress difference, the apparent dynamic viscosity can be computed by $\eta = \tau_{xy}/\dot\gamma$. The shear strain rate dependent relaxation time of polymer solution is calculated from $\lambda = N_1/(2\dot\gamma^2\eta_p)$ with $\eta_p$ being the polymer viscosity.

At the macroscale level, the bulk rheology of the polymer solution is modeled by the continuum equations given by Eqs.~\eqref{eq:Gov}. We consider a 0.5\% aqueous polyacrylamide solution according to Purnode's experiments~\cite{1996Purnode}. We take the dynamic viscosity $\eta_0 = 0.378~{\rm Pa\,s}$ at a shear strain rate of $\dot\gamma_0 = 0.504~{\rm s^{-1}}$ in the experimental data as the zero shear viscosity, and the normal stress difference at a median shear strain rate $\dot\gamma = 17.625~{\rm s^{-1}}$ is $N_1 =3.925~{\rm Pa}$. Based on these physical quantities, we can determine the length unit~$[L]=1.44\times10^{-4}~{\rm m}$, the mass unit~$[M]=9.99\times10^{-10}~{\rm kg}$, and the time unit~$[T]=1.1\times10^{-3}~{\rm s}$ of the DPD system. For hydrodynamic and rheology problems, all the physical quantities with physical units can be mapped to reduced DPD units by dimensional analysis using the three basic units $[L]$, $[M]$ and $[T]$. For example, the shear strain rate is converted by the shear strain rate unit $[\dot\gamma]=1/[T]=903.22~{\rm s}^{-1}$, the dynamic viscosity is converted by the viscosity unit $[\eta]=[M][L]^{-1}[T]^{-1}=6.304\times10^{-3}~{\rm Pa\,s}$, the shear stress is converted by the stress unit $[\tau]=[M][L]^{-1}[T]^{-2}=5.694~{\rm Pa}$, and the first normal stress difference is $[N_1]=2[M][L]^{-1}[T]^{-2}=11.39~{\rm Pa}$.

\section{Numerical Implementation and Results}\label{sec:3}
We first simulate a start-up viscoelastic flow up to steady state in a parallel-plate channel to demonstrate the proposed framework of employing the active-learning algorithm applied to multiscale simulation of viscoelastic fluids; subsequently, we simulate a viscoelastic flow past a circular cylinder to demonstrate how transfer-learning works in the micro-macro scales coupling. In both cases, the constitutive closure models used in the continuum equations are
constructed on-the-fly from microscopic simulations.

For the pressure-driven start-up viscoelastic flow, we do not have a priori the flow velocity in steady state. In the DPD model of polymer solution, the zero-shear dynamic viscosity of the polymer solution is $\eta_0 = 59.95$ in reduced DPD units and the DPD solvent has a dynamic viscosity of $\eta_s = 1.40$, hence the viscosity ratio is $\beta = \eta_s/\eta_0 = 2.34\times10^{-2}$. Given a characteristic kinematic viscosity $\tilde{\nu}=1.76\times10^{-4}~m^2/s$, we consider a polymer solution with mass density $\rho = 1000~kg/m^3$, solvent viscosity $\nu_s = 0.05\tilde{\nu}$ and polymer viscosity at zero-shear-rate $\nu_p = 2.09\tilde{\nu}$. The viscosity of the polymer solution is $\nu_0 = \nu_s + \nu_p = 2.14\tilde{\nu}$ with a viscosity ratio $\beta = \eta_s/\eta_0 = \nu_s/\nu_0 = 2.34\times10^{-2}$. Also, the zero-shear-rate relaxation time is $\lambda_0 = 1.94~s$.
To specify the input to the SEM solver, we introduce a prior velocity scale denoted by $\tilde{U}$, based on which we define two dimensionless numbers $\tilde{R}=\tilde{U}L_0/\tilde{\nu}$ and $\tilde{W}=\lambda_0\tilde{U}/L_0$. Correspondingly, the dimensionless driving force can be determined by $\tilde{f}= \nabla P\cdot 2L_0/(\rho\tilde{U}^2)$.
We will use the following four examples including two channel flows and two flows past a cylinder confined in a channel to demonstrate the active- and transfer-learning methodology, as listed in Table~\ref{Table_1}.

\begin{table}[h!]
\centering
\caption{System setup of the parameters $\widetilde{R}$, $\widetilde{W}$, $\widetilde{f}$, $L_0$, $\widetilde{U}$ and $f=\nabla P$ for the four examples}
\label{Table_1}
\begin{tabular}{ccccccc}
\hline\hline\vspace{-0.39cm}\\
\ & $\widetilde{R}$ & $\widetilde{W}$ & $\widetilde{f}$ & $L_0[cm]$ & $\widetilde{U}[cm/s]$ & $f=\nabla P[kg/(m^2s^2)]$\\
\hline
Channel Flow \#1 & 10 & 1 & 2 & 5.85 & 3.01 & 15.49\\
Channel Flow \#2 & 1 & 5 & 4 & 0.83 & 2.13 & 109.32\\
Flow Past Cylinder \#1 & 1 & 2 & 3 & 1.31 & 1.35 & 20.87\\
Flow Past Cylinder \#2 & 20 & 1 & 0.4 & 8.27 & 4.26 & 4.39\\
\hline\hline
\end{tabular}
\end{table}

In the following, we will show how the GPR model, constructed in one channel flow, can be transferred to another channel flow with different operating conditions. Also, we will likewise demonstrate how the GPR model constructed in channel flows can be transferred to flow past a circular cylinder by actively initializing a few more DPD simulations.

\subsection{Parallel-plate Channel Flows}\label{sec:3_1}
Flow between two parallel plates, as shown in Fig.~\ref{FIG:figure1}, is used to demonstrate the application of the active-learning scheme to microscale-macroscale coupling for modeling viscoelastic flows. At the macroscopic level, the bulk rheology of the polymer solution is modeled by the continuum equations in the form of Eq.~\eqref{eq:Gov}, which are discretized and solved by SEM~\cite{2003Ma} with an entropy-viscosity method (EVM)~\cite{2019Wang} for numerical stabilization at high Weissenberg numbers. Let coordinates $x$ and $y$ denote the streamwise and transverse (wall-normal) directions, and $u$ and $v$ represent the corresponding velocity components, respectively. We perform the SEM simulation in a two-dimensional rectangular computational region of $20L_0\times2L_0$, with two flat plates fixed at $y=\pm L_0$ as the wall boundaries, as shown in Fig.~\ref{FIG:figure1}. The computational domain is discretized by $8\times6$ spectral elements with a polynomial order of $p=7$. The inlet is at $x=-10L_0$, the outlet is at $x=10L_0$.

The channel flow is initialized with a zero velocity field. A pressure gradient is applied to drive the polymer solution to flow through the channel. The periodic boundary condition is applied in the $x$-direction, while the surfaces of the upper and lower plates are set to be solid walls with a no-slip ($u=0, v=0$) boundary condition.
The Reynolds number is defined by $Re = UL/\nu_0 = 2\tilde{R}U/(2.14\tilde{U})$ with the channel width $L = 2L_0$, the flow velocity $U$ and the zero-shear-rate viscosity $\nu_0=2.14\tilde{\nu}$. Flows at different $Re$ can be generated by applying various pressure gradients to drive the flow, or by changing the zero-shear-rate viscosity of the fluid. The elasticity of the polymer solution is described by the Weissenberg number, which is defined as the ratio of the microscopic time scale to the macroscopic strain rate, i.e., $We = \lambda_0 U/L = 0.5\tilde{W}U/\tilde{U}$.

In the microscale-macroscale coupled simulation, a FENE-P model, modified to have shear-thinning polymer viscosity and relaxation time, is used to close the continuum equations. They are discretized and solved by SEM on rectangular elements, while the microscale dynamics of polymer chains is simulated with DPD. The continuum solver acts as a master solver which actively calls the slave DPD simulator on-the-fly as necessary. A GPR model is employed to decide when and where to do the next sampling. To initialize the GPR model for active-learning, we first set up the SEM system and perform three DPD simulations of simple shear flow at low strain rates. The effective dynamic viscosity and relaxation time of the polymer solution can be derived from the computed stress tensor by $\eta = \tau_{xy}/\dot\gamma$ and $\lambda = (\tau_{xx}-\tau_{yy})/(2\dot\gamma^2\eta_p)$.
Three DPD simulations of the polymer solution in simple shear flow are performed to generate the initial training data for GPR, namely at strain rates in reduced units: $\dot\gamma = 0.5~{\rm s^{-1}}$ ($5.5\times10^{-4}$ in reduced unit), $\dot\gamma = 1.0~{\rm s^{-1}}$ ($1.1\times10^{-3}$ in reduced unit) and $\dot\gamma = 1.5~{\rm s^{-1}}$ ($1.65\times10^{-3}$ in reduced unit).

The GPR model predicts not only the mean apparent viscosity $\eta(\dot\gamma)$ and normal stress difference $N_1(\dot\gamma)$ at various strain rates, but also their corresponding uncertainties in terms of standard deviations $\sigma_\eta(\dot\gamma)$ and $\sigma_{N_1}(\dot\gamma)$. Because the range of the shear strain rate during the transient channel flow could change by several orders of magnitude, we perform the GPR on logarithmic scales, e.g., $\log\eta$ and $\log N_1$ versus $\log\dot\gamma$. Given the GPR prediction uncertainties, we can define an acquisition function $\alpha_\eta = \sigma_\eta(\dot\gamma)$ for viscosity $\eta$ and an acquisition function $\alpha_{N_1} = \sigma_{N_1}(\dot\gamma)$ for $N_1$ to indicate the location of next optimal sampling point. We note that the acquisition function $\alpha$ can also be defined by other functions of prediction uncertainty as well as its magnitude and curvature, wherein the selection of acquisition function would depend on the intrinsic smoothness of target functions in different problems. Let $\alpha = \max\{\alpha_\eta, \alpha_{N_1}\}$ represent the maximum prediction uncertainty, we choose an acceptance criterion $\max(\alpha)\le\delta_{\rm tol}$ for the GPR-informed effective viscosity and normal stress difference, where $\delta_{\rm tol} = 1\%$ is a predefined tolerance.

\begin{figure}[h!]
\centering
\includegraphics[width=0.6\textwidth]{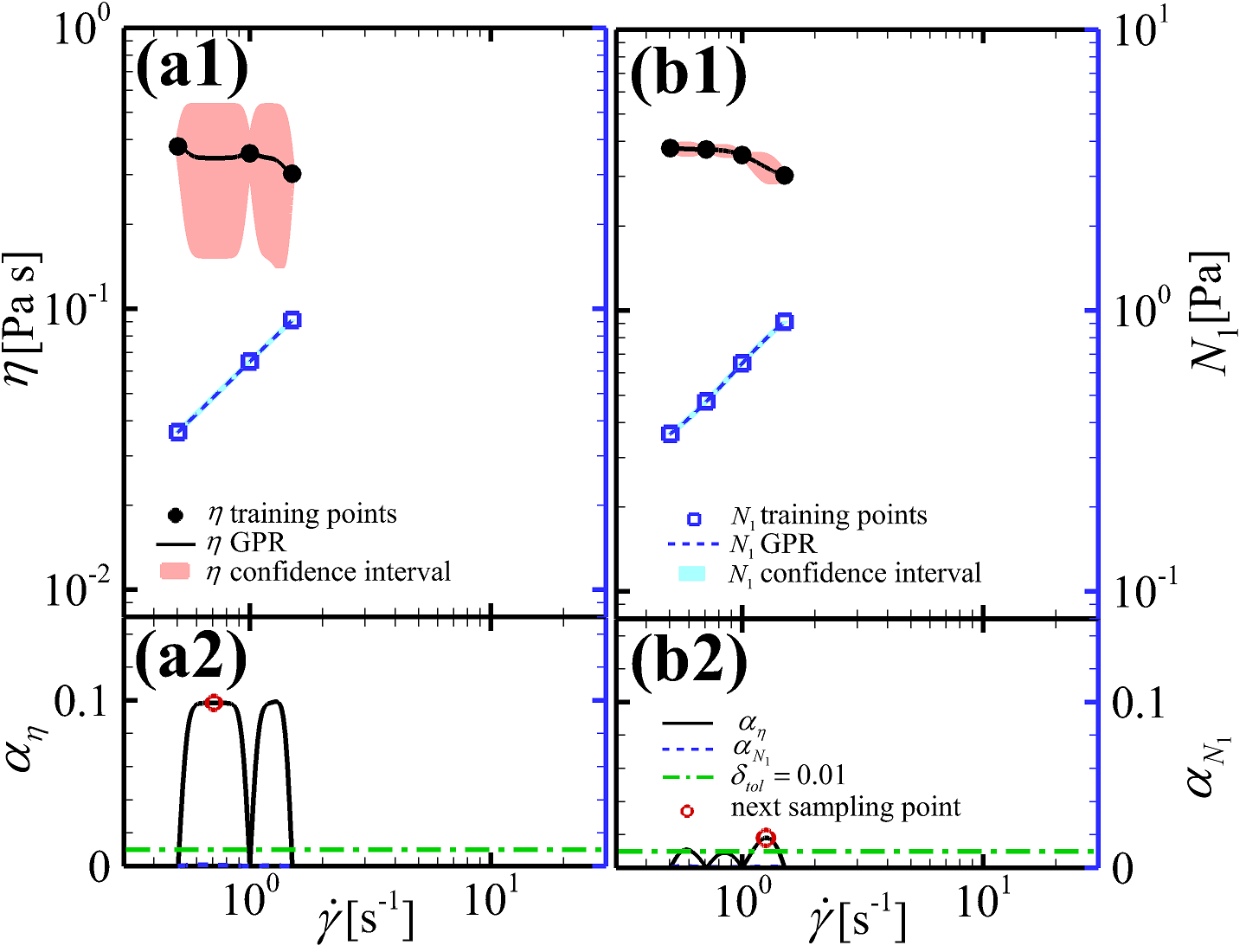}
\caption{Active-learning of viscosity, $\eta$, and normal stress difference, $N_1$, based on GPR and using (a1) three training points, and (b1) four training points obtained from DPD simulations. The training data are shown as filled circles for $\eta$, and open squares for $N_1$. The  GPR predictions are shown as lines, i.e., solid for $\eta$ and dotted for $N_1$. The uncertainties are visualized as the shaded areas bounded by the $95\%$ confidence level. The plots (a2) and (b2) show the acquisition functions $\alpha_\eta(\dot\gamma)$ and $\alpha_{N_1}(\dot\gamma)$, respectively. Their maxima indicate the next sampling points $\dot\gamma_{p1}=0.71~s^{-1}$ in (a2) and $\dot\gamma_{p2}=1.26~s^{-1}$ in (b2) for interpolation.}
\label{FIG:figure3}
\end{figure}

The results of DPD simulation performed at three low shear strain rates ($\dot\gamma_{O_1} = 0.5~{\rm s^{-1}}$, $\dot\gamma_{O_2} = 1~{\rm s^{-1}}$ and $\dot\gamma_{O_3} = 1.5~{\rm s^{-1}}$) are taken as the initial training points of a GPR model. Fig.~\ref{FIG:figure3}(a1) presents the GPR predictions of dynamical viscosity $\eta$ and normal stress difference $N_1$ based on the three training points, where the solid line is the GPR prediction of $\eta$ and the dotted line is the prediction of $N_1$, and the training data for $\eta$ and $N_1$ are represented by circles and squares, respectively. The predicted  uncertainties are visualized by the $95\%$ confidence interval shown as the shaded area. Fig.~\ref{FIG:figure3}(a2) plots the magnitude of the corresponding acquisition function $\alpha_\eta(\dot\gamma)$ and $\alpha_{N_1}(\dot\gamma)$, where the predefined tolerance $\delta_{\rm tol} = 1\%$ is shown by the horizontal dot-dashed line. It is obvious in Fig.~\ref{FIG:figure3}(a2) that the magnitude of the acquisition function $\alpha = \max\{\alpha_\eta, \alpha_{N_1}\}$ reaches its maximum value $\alpha = 0.10$ at the shear strain rate $\dot\gamma = 0.71~{\rm s^{-1}}$, which is far beyond the designed tolerance $\delta_{\rm tol}$ and breaks the acceptance criterion. Therefore, we need additional training points by running DPD simulations to reduce the magnitude of $\alpha$. In practice, we take the shear strain rate $\dot\gamma_{p_1}$ at the maximum $\alpha$ as the next sampling point. Then, we perform a simple shear DPD simulation of the polymer solution at $\dot\gamma_{p_1}=0.71~{\rm s^{-1}}$ and obtain the corresponding $\eta$ and $N_1$ derived from the computed stress tensor. With only one additional data point, we observe a better GPR prediction of $\eta$ and $N_1$ with smaller uncertainties as shown in Fig.~\ref{FIG:figure3}(b1). Although the magnitude of $\alpha$ is significantly reduced by adding one data point at $\dot\gamma_{p_1}$, the maximum of $\alpha$ is still larger than the tolerance $\delta_{\rm tol}$. Similarly, we need to perform another simple shear DPD simulation at the next sampling point $\dot\gamma_{p_2} = 1.26~{\rm s^{-1}}$, which is indicated by the location of maximum $\alpha$ in Fig.~\ref{FIG:figure3}(b2). This process of adding new points can be repeated until the GPR predicted uncertainties in terms of $\alpha$ are smaller than the prescribed tolerance $\delta_{\rm tol}$. Fig.~\ref{FIG:figure5}(a1) presents the GPR predictions of $\eta$ and $N_1$ based on the five data points, whose uncertainty described by $\alpha$ is smaller than the tolerance $\delta_{\rm tol}$ over the range of shear strain rate up to $\dot\gamma=1.86~{\rm s^{-1}}$.

Because the flow field of the SEM system is initialized with zero velocity, we can advance the time integration of the SEM solver using the GPR predictions of $\eta$ and $N_1$. The GPR model based on the five data points shown in Fig.~\ref{FIG:figure5}(a1) is valid until the global maximum of $\dot\gamma_{\max}$ in the SEM simulation exceeds a value that breaks the acceptance criterion $\max(\alpha)\le\delta_{\rm tol}$. In the start-up channel flow, the global maximum of the strain rate $\dot\gamma_{\max}$ in the SEM system changes with time before the flow reaches a steady state. Therefore, we need to monitor the magnitude of the acquisition function $\alpha(\dot\gamma)$ (up to $\dot\gamma_{\max}$) in order to decide when we need to run additional DPD simulations. We can see in Fig.~\ref{FIG:figure5}(a2) that the acquisition functions $\alpha_{\eta}$ and $\alpha_{N_1}$ increase sharply with the shear strain rate $\dot\gamma$. Let $\dot\gamma_{\rm gpr}$ be the maximum strain rate covered by a valid GPR-prediction $\eta$ and $N_1$ with $\alpha \le \delta_{\rm tol}$. When the global maximum of strain rate $\dot\gamma_{\max}$ in the channel flow is larger than $\dot\gamma_{\rm gpr}$, we need additional training points for extrapolation to reduce the magnitude of $\alpha$. Unlike interpolation of new points in GPR, the extrapolation of the next sampling point $\dot\gamma_p$ requires the size of the jump to be determined. To that end, a tunable parameter $\beta$ is introduced to determine jump size for extrapolation, so that the extrapolation point is found from $\alpha(\dot\gamma_p) = \beta\delta_{\rm tol}$. In general, too small a $\beta$ will fail to reduce the number of training points while a too large one may break the acceptance criterion, and require extra interpolating points. The optimum choice for $\beta$ achieves $\max(\alpha)\simeq\delta_{\rm tol}$ with only a single extrapolation sampling point.

\begin{figure}[t]
\centering
\includegraphics[width=0.5\textwidth]{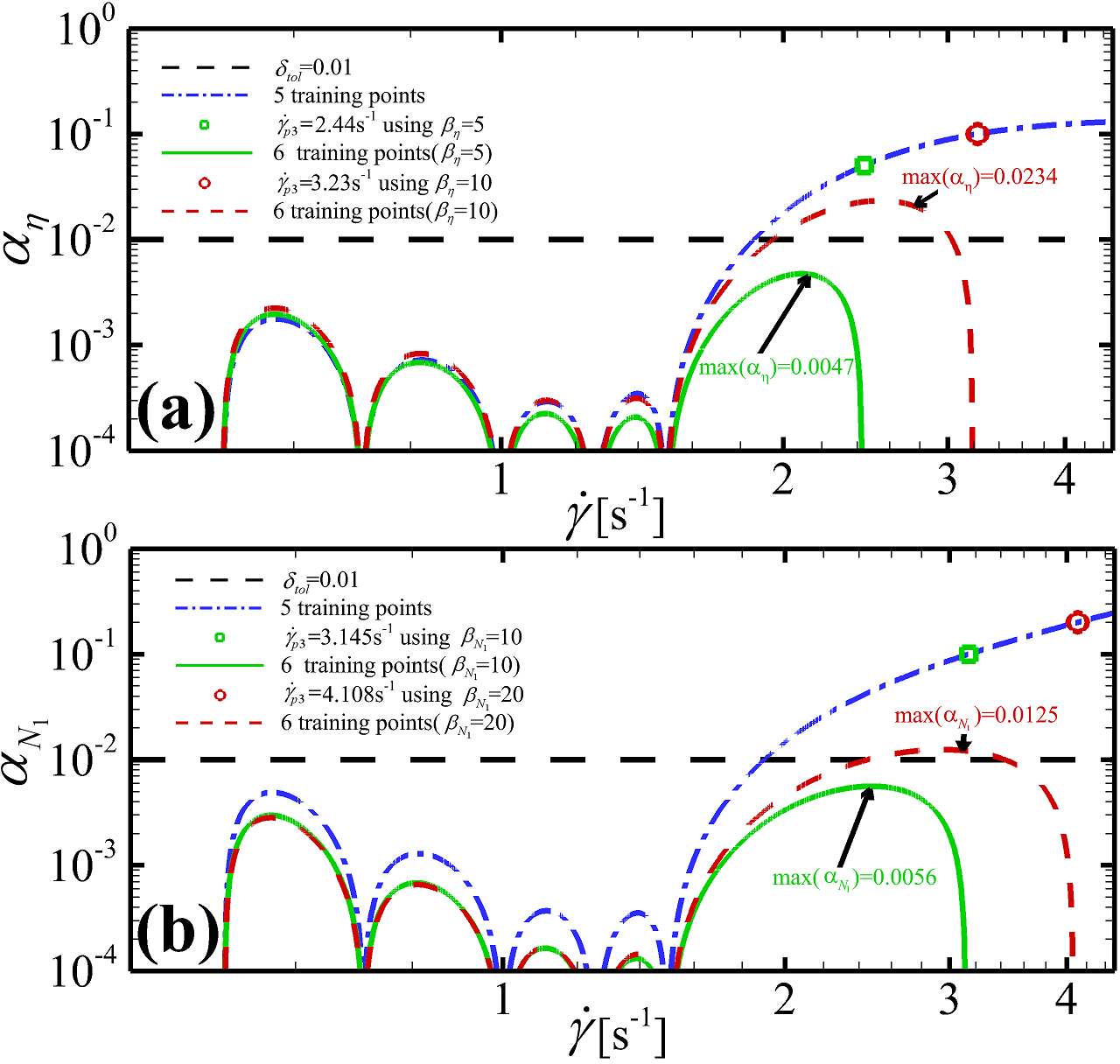}
\caption{Discovering the appropriate extrapolation step-size in active sampling using the acquisition function $\alpha_\eta$. (a) Its magnitude after adding one extrapolation point using $\beta_\eta=5$ and $\beta_\eta=10$, and (b) its magnitude after adding one extrapolation point using $\beta_{N_1}=10$ and $\beta_{N_1}=20$. The horizontal dashed lines in (a) and (b) denote the acceptance criteria for a predefined tolerance $\delta_{\rm tol}=0.01$. Sampling with $\beta_\eta = 10$ is too aggressive for extrapolation of ${\eta}(\dot\gamma)$ for a tolerance of $0.01$, and $\beta_{N_1} = 20$ is likewise too large for extrapolation of $N_1(\dot\gamma)$.}
\label{FIG:figure4}
\end{figure}
\begin{figure}[t]
\centering
\includegraphics[width=0.8\textwidth]{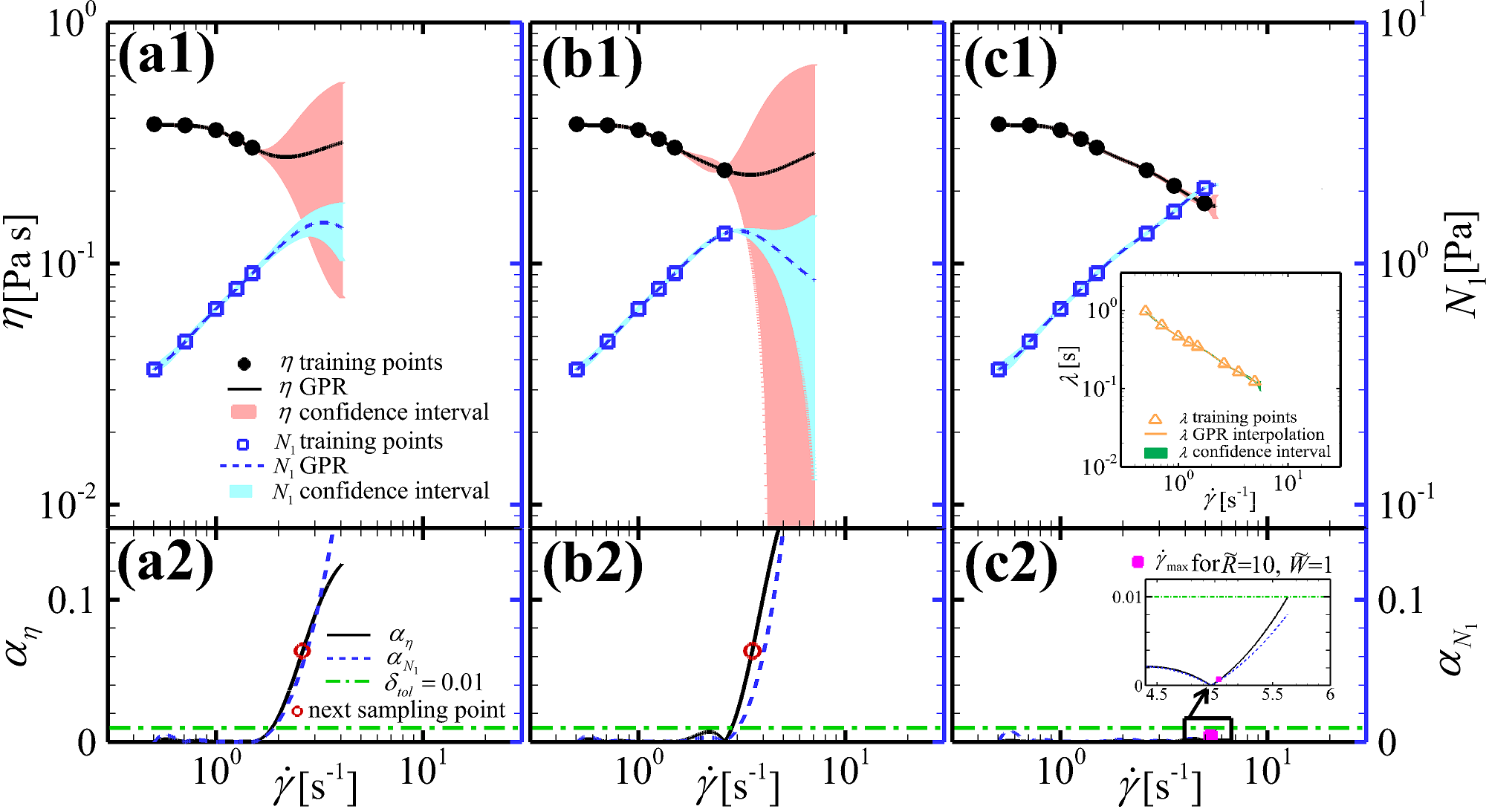}
\caption{Determination of material functions $\eta$ and $N_1$ by extrapolation using GPR: with (a1) five training points and with (b1) six training points obtained from DPD simulations. Filled circles and open squares represent training data for $\eta$ and $N_1$ respectively, GPR predictions for
$\eta$ and $N_1$ are shown as solid and dotted lines, respectively. Uncertainties are visualized
as the shaded areas bounded by the $95\%$ confidence interval. Extrapolation of the next sampling to be $\dot\gamma_{p3}=2.61~s^{-1}$ in (a2) and $\dot\gamma_{p4}=3.55~s^{-1}$ in (b2) obtained from $\alpha_\eta= \beta_\eta \delta_{\rm tol}$, $\beta_\eta = 6.41$. (c1) shows the macroscale-microscale coupled system is eventually closed with 7 sampling points in total, with an inset showing the dependence of relaxation time $\lambda$ on the strain rate $\dot\gamma$.}
\label{FIG:figure5}
\end{figure}

To find a proper value of $\beta$ for extrapolation in the GPR, we perform a few numerical experiments using different values of $\beta$ as shown in Fig.~\ref{FIG:figure4}. In the GPR of the effective viscosity $\eta$, we first try $\beta_\eta = 5$ and obtain the next sampling point at $\dot\gamma_{p_3} = 2.44~{\rm s^{-1}}$ by setting $\alpha(\dot\gamma_{p_3}) = 5\delta_{\rm tol}$. After adding one sampling point at $\dot\gamma_{p_3} = 2.44~{\rm s^{-1}}$, we observe in Fig.~\ref{FIG:figure4}(a) that $\max(\alpha) = 0.47\% < \delta_{\rm tol}/2$, which means that $\beta_{\eta} = 5$ results in an overcautious jump for extrapolation. Subsequently, we try a larger value of $\beta_\eta = 10$, leading to the next sampling point at $\dot\gamma_{p_3} = 3.23~{\rm s^{-1}}$. This leads to $\max(\alpha) = 2.34\% > \delta_{\rm tol}$ and breaks the acceptance criterion, as shown in Fig.~\ref{FIG:figure4}(a). Consequently, using $\beta_\eta = 10$ may be too aggressive for extrapolation and we will need to add extra interpolating points to satisfy $\max(\alpha)|_{\dot\gamma < \dot\gamma_{p_3}} < \delta_{\rm tol}$. Based on the tests on $\beta_\eta = 5$ and $\beta_\eta = 10$ in Fig.~\ref{FIG:figure4}(a), using $85\%$ of the value obtained by a linear interpolation, we estimate that $\beta_\eta = 6.41$ could be a good option for extrapolation with the prescribed tolerance $\delta_{\rm tol} = 1\%$.
Using the same method, we test $\beta_{N_1} = 10$ and $\beta_{N_1} = 20$ for the GPR of the first normal stress difference $N_1$. We observe in Fig.~\ref{FIG:figure4}(b) that $\beta_{N_1} = 10$ results in $\max(\alpha) = 0.56\%$ and $\beta_{N_1} = 20$ results in $\max(\alpha) = 1.25\%$, from which an optimal value for extrapolation of $N_1$ is $\beta_{N_1} = 14.19$.
To reduce the number of DPD simulations, we selected a proper value of $\beta$ to avoid performing extra interpolating samplings after each extrapolation, with the next sampling point determined by
\begin{equation}\label{eq:beta}
  \dot\gamma_p = \min(\dot\gamma_{\eta,p}, \dot\gamma_{N_1,p}),
\end{equation}
where the magnitudes of $\dot\gamma_{\eta,p}$ and $\dot\gamma_{N_1,p}$ are computed by $\alpha_\eta(\dot\gamma_{\eta,p}) = \beta_\eta \delta_{\rm tol}$ and $\alpha_{N_1}(\dot\gamma_{N_1,p}) = \beta_{N_1} \delta_{\rm tol}$, respectively.
Fig.~\ref{FIG:figure5} shows the extrapolating process in the GPR model to construct the effective viscosity $\eta$ and the first normal stress difference $N_1$. More specifically, Fig.~\ref{FIG:figure5}(a2) and (b2) plot the acquisition functions $\alpha_\eta(\dot\gamma)$ and $\alpha_{N_1}(\dot\gamma)$, in which the next sampling points $\dot\gamma_{p3}=2.61~s^{-1}$ in Fig.~\ref{FIG:figure5}(a2) and $\dot\gamma_{p4}=3.55~s^{-1}$ in Fig.~\ref{FIG:figure5}(b2) are determined by $\alpha(\dot\gamma) = \beta_\eta \delta_{\rm tol}$ using $\beta_\eta = 6.41$. Fig.~\ref{FIG:figure5}(c1) shows that the macroscale-microscale coupled system of a channel flow at $\tilde{R}=10$ and $\tilde{W}=1$ is eventually closed with 7 sampling points in total, and its inset shows the learned relaxation time $\lambda$ as a function of shear strain rate $\dot\gamma$. Fig.~\ref{FIG:figure_channel}(a) plots the distribution of the shear strain rate $\dot{\gamma}$, the dynamic viscosity $\eta$ and relaxation time $\lambda$ in the channel flow at $\tilde{R}=10$ and $\tilde{W}=1$, showing that both the effective viscosity $\eta$ and the relaxation time $\lambda$ are decreased near the wall surfaces because of increased shear strain rate $\dot\gamma$. The inset of Fig.~\ref{FIG:figure_channel}(a) shows that the global maximum of strain rate in the channel flow increases with time before reaching $\dot{\gamma}_{\max}=3.20~{\rm s^{-1}}$ at the steady state, while the GPR model for $\eta$ and $N_1$ based on seven training points is valid up to $\dot{\gamma}_{\rm gpr}=3.93~{\rm s^{-1}}$. Consequently, the GPR-informed $\eta$ and $N_1$ computed from the microscopic dynamics are sufficient to close the macroscopic equations with the modified FENE-P model in this channel flow.

\begin{figure}[t!]
\centering
\includegraphics[width=0.9\textwidth]{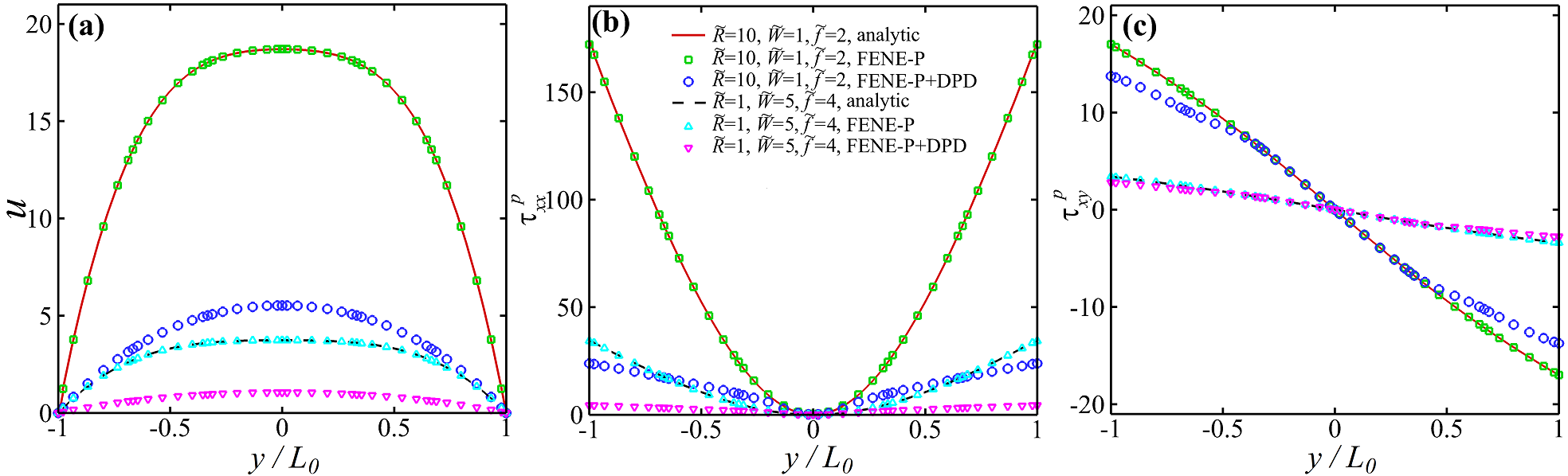}
\caption{Channel flow profiles: (a) velocity $u(y)$, (b) polymer normal stress $\tau^p_{xx}(y)$ and (c) polymer shear stress $\tau^p_{xy}(y)$, under operating conditions $\tilde{R}=10,~\tilde{W}=1,~\tilde{f}=2$ and $\tilde{R}=1,~\tilde{W}=5,~\tilde{f}=4$ for the FENE-P model (FENE-P) solved by analytic and SEM methods, and for the coupled macroscale-microscale solutions (FENE-P+DPD).}
\label{FIG:figure8}
\end{figure}

Using the same driving force, we performed simulations using the classical FENE-P model (FENE-P) and also the macroscale-microscale coupled model (FENE-P+DPD).
Fig.~\ref{FIG:figure8} shows different profiles of the polymer solutions flowing through a channel, including the velocity profile $u(y)$ in Fig.~\ref{FIG:figure8}(a), the polymer normal stress profile $\tau^p_{xx}(y)$ in Fig.~\ref{FIG:figure8}(b), and the polymer shear stress profile $\tau^p_{xy}(y)$ in Fig.~\ref{FIG:figure8}(c). For the channel flow with $\tilde{R}=10$ and $\tilde{W}=1$, our SEM solutions for a classical FENE-P model with constant parameters are represented by the squares, which is consistent with the analytic solution of a FENE-P model (solid line) and verifies the SEM solver. The modified FENE-P model with shear-thinning viscosities $\eta(\dot\gamma)$ and relaxation times $\lambda(\dot\gamma)$ results in the macroscale-microscale coupling solutions represented by circles. It is obvious in Fig.~\ref{FIG:figure8}(a) that the FENE-P+DPD model has a smaller flow rate than the classical FENE-P model in the channel flow. The FENE-P+DPD model has smaller viscosity than the classical FENE-P model because of shear thinning effects, as shown in Fig.~\ref{FIG:figure5}(c1); the observation in Fig.~\ref{FIG:figure8}(a) would contradict the physical intuition about inelastic fluids that a less viscous fluid should have a higher flow rate in the channel flow. After the flow reaches its steady state, the maximum flow velocity in the classical FENE-P model is $U = 56.31~cm/s$ ($18.71$ in reduced unit) resulting in $Re=174.29$ and $We = 9.34$, while the flow velocity in the macroscale-microscale coupled model is $U = 16.63~cm/s$ ($5.52$ in reduced unit) resulting in $Re = 51.47$ and $We=2.76$.

\begin{figure}[t!]
\centering
\includegraphics[width=0.8\textwidth]{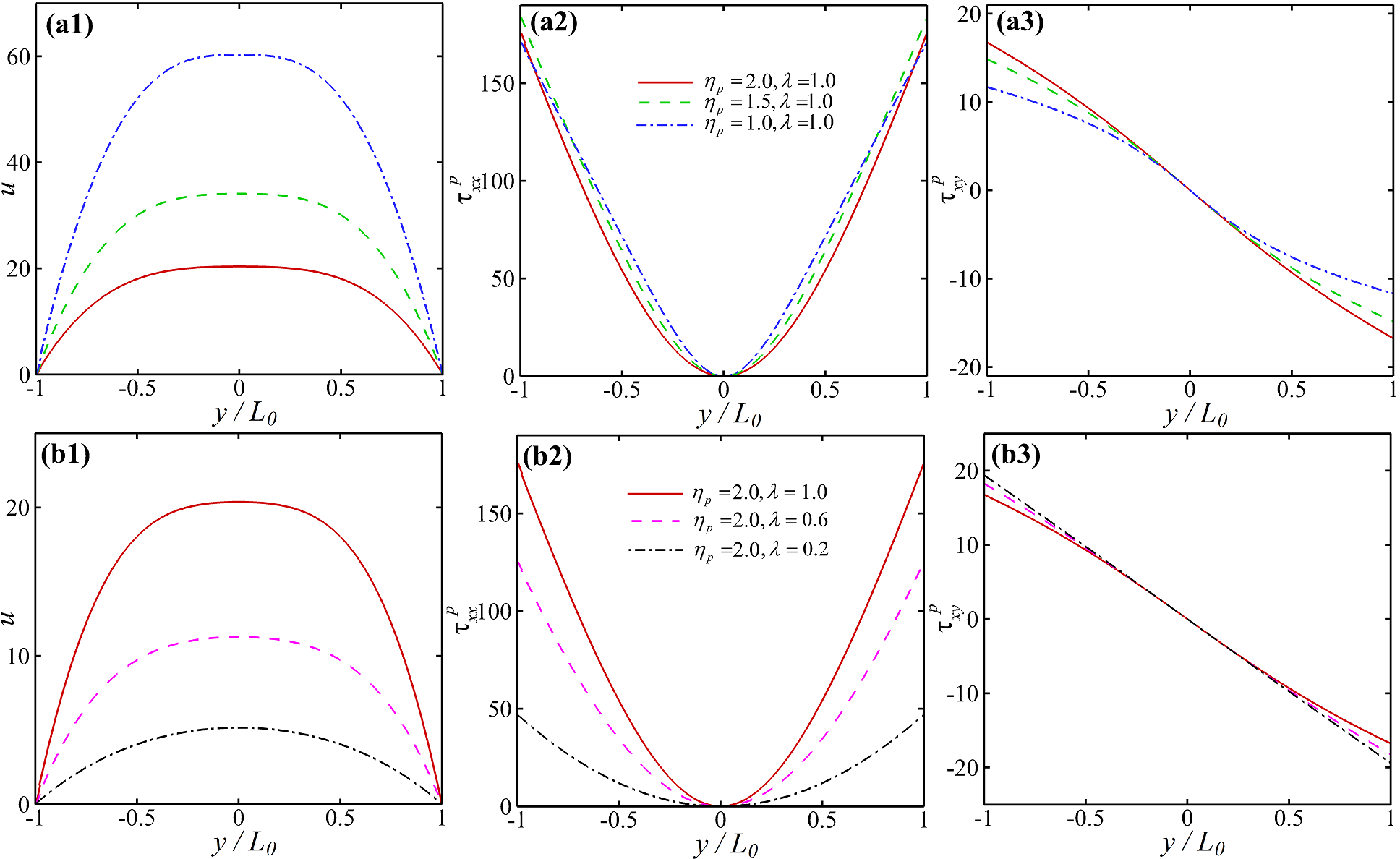}
\caption{Competing trends determine the velocity profiles. Parametric sensitivity analysis of FENE-P model of channel flow for velocity $u(y)$, polymer normal stress $\tau^p_{xx}(y)$ and polymer shear stress $\tau^p_{xy}(y)$ at (a1-a3) $\lambda = 1.0,~\eta_p = 1.0,~1.5,~2.0$ and (b1-b3) $\eta_p = 2.0,~\lambda = 1.0,~0.6,~0.2$.}
\label{FIG:figure9}
\end{figure}

\begin{figure}[t!]
\centering
\includegraphics[width=0.8\textwidth]{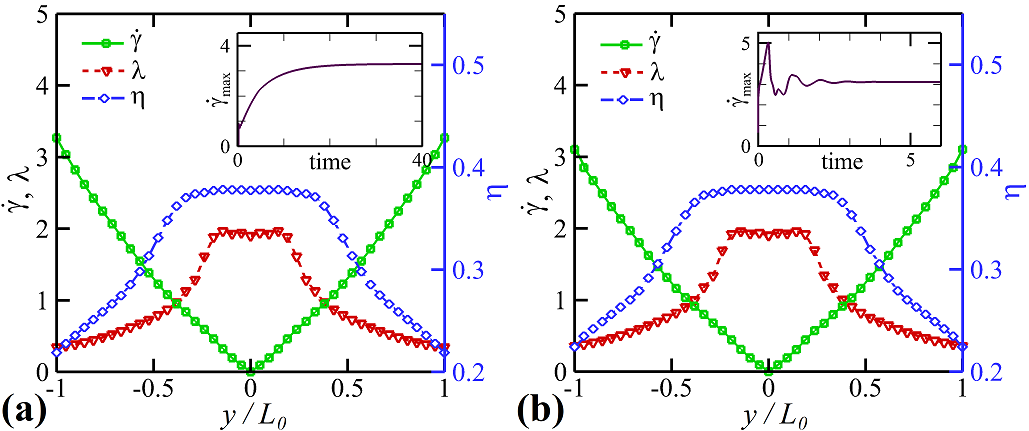}
\caption{Profiles of the shear strain rate $\dot{\gamma}$, the dynamic viscosity $\eta$ and the relaxation time $\lambda$ in (a) Channel flow \#1 at $\tilde{R}=10$ and $\tilde{W}=1$, and (b) Channel flow \#2 at $\tilde{R}=1$ and $\tilde{W}=5$ after the flows reach their steady states. The insets show the time evolution of the maximum shear strain rate $\dot{\gamma}_{\rm max}$ in the channel flows.}
\label{FIG:figure_channel}
\end{figure}

With straight path lines steady channel flow velocity profiles are determined solely by the viscosity function. Hence the fluids elasticity affects the flow only in so far as a change of relaxation time alters the viscosity function.
Fig.~\ref{FIG:figure9} shows a parametric sensitivity analysis of the FENE-P model in a channel flow, affecting the velocity profile $u(y)$, the polymer normal stress profile $\tau^p_{xx}(y)$ and the polymer shear stress profile $\tau^p_{xy}(y)$. In Fig.~\ref{FIG:figure9}(a1-a3), we decrease the polymer viscosity $\eta_p$ from $2.0$ to $1.0$ while keeping the relaxation time $\lambda$ constant. We observe in Fig.~\ref{FIG:figure9}(a1) that the FENE-P fluid with a decreased viscosity has a higher flow rate, which is consistent with the physical intuition of viscous flows. However, when we decrease the relaxation time $\lambda$ from $1.0$ to $0.2$ while keeping the polymer viscosity $\eta_p$ constant in Fig.~\ref{FIG:figure9}(b1-b3), we find that the FENE-P fluid with a decreased relaxation time has a lower velocity profile and a decreased flow rate, which is an opposite trend as the effects of decreasing the viscosity. Therefore, the flow profiles of the microscale-macroscale coupled simulation (FENE-P+DPD) shown in Fig.~\ref{FIG:figure8} are determined by the competition between viscous effects and elastic effects in the channel flow.

\begin{figure}[t!]
\centering
\includegraphics[width=0.6\textwidth]{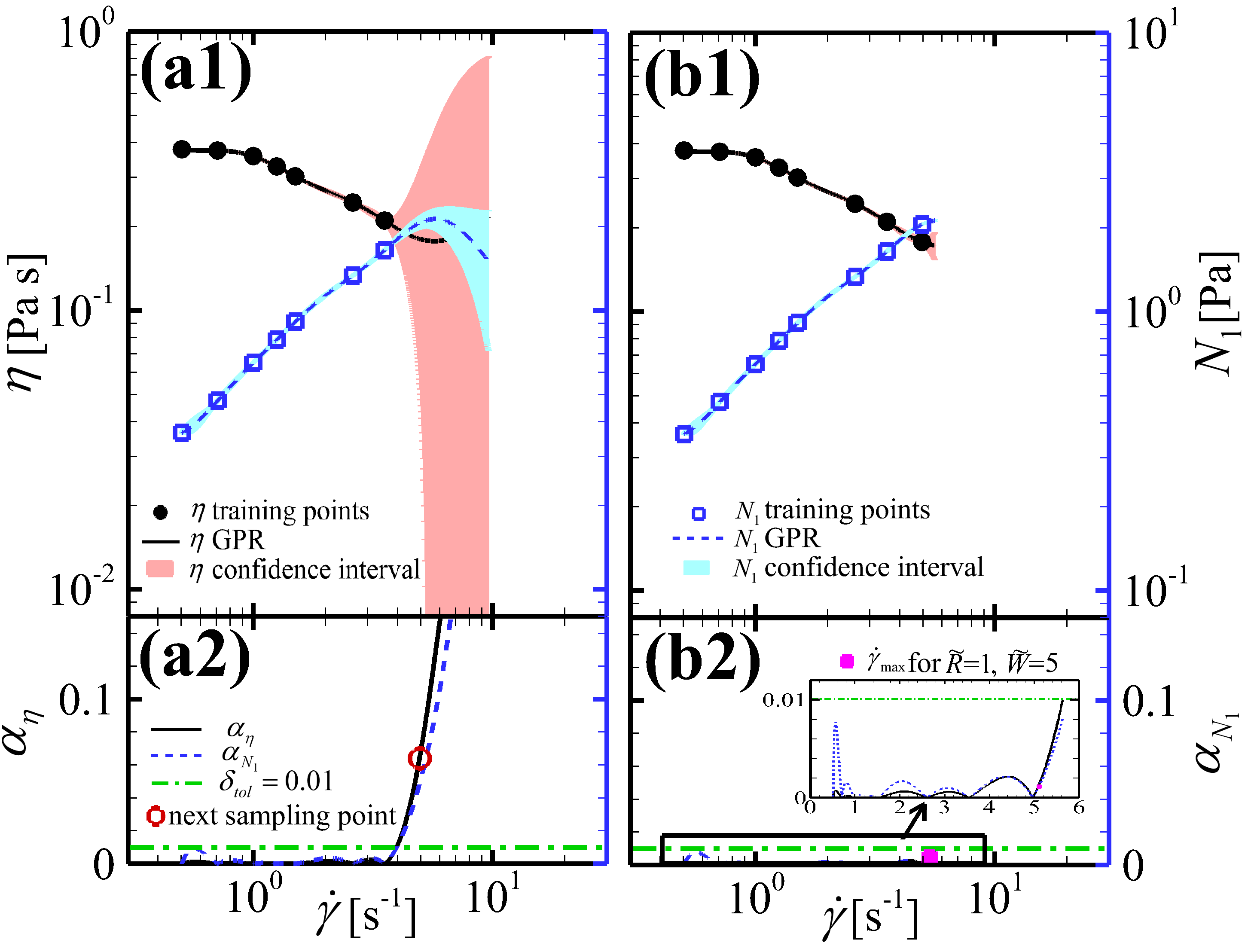}
\caption{Transfer-learning from channel \#1 ($\tilde{R}=10$ and $\tilde{W}=1$) to channel \#2 ($\tilde{R}=1$ and $\tilde{W}=5$). (a) Previously learned material functions in channel \#1, and (b1) GPR-informed material functions based on previous results with one additional sampling point. Training data: for $\eta$, filled circles, data for $N_1$, squares. GPR predictions: for $\eta$, solid lines, and $N_1$, dotted lines. Uncertainties are visualized as the shaded areas bounded by the $95\%$ confidence level. The additional extrapolation sampling-point $\dot\gamma_{p5}=4.97~s^{-1}$ is determined from $\alpha_\eta(\dot\gamma_{p5}) = \beta_\eta\delta_{\rm tol}$, $\beta_\eta = 6.41$. The filled square in (b2) is the global maximum strain-rate $\dot\gamma_{\max} = 5.08~s^{-1}$ at $\tilde{R}=1$ and $\tilde{W}=5$.}
\label{FIG:figure6}
\end{figure}

For a shear-thinning fluid flowing though a straight channel, a higher Weissenberg number can yield a more flattened velocity profile around the channel centerline and a sharper layer in the vicinity of the wall, which will increase the maximum strain rate in the system. For simplicity, we call the case of channel flow at $\tilde{R}=10$ and $\tilde{W}=1$ as channel \#1, and another case of channel flow at $\tilde{R}=1$ and $\tilde{W}=5$ as channel \#2. The GPR model for the effective viscosity $\eta$ and the first normal stress difference $N_1$ (computed for the relaxation time) constructed in the channel \#1 can be transferred to the channel \#2. As shown in Fig.~\ref{FIG:figure5}(c2). the GPR model for $\eta$ and $N_1$ based on the seven training points in the channel \#1 is valid up to $\dot\gamma_{\rm gpr} = 3.93~{\rm s^{-1}}$. Fig.~\ref{FIG:figure_channel}(b) plots the profiles of the shear strain rate $\dot{\gamma}$, the dynamic viscosity $\eta$ and relaxation time $\lambda$ in channel flow \#2. The inset of Fig.~\ref{FIG:figure_channel}(b) shows the time evolution of the global maximum strain rate of $\dot\gamma_{\max}$, where an overshoot response induced by the increased $\lambda$ is observed, leading to $\dot\gamma_{\max}=5.08~{\rm s^{-1}}$ beyond the point $\dot\gamma_{\rm gpr} = 3.93~{\rm s^{-1}}$ in Fig.~\ref{FIG:figure5}(c2). Then, we need to perform additional DPD simulations to include new data points once the acquisition function $\alpha$ breaks the acceptance criterion $\max(\alpha)<\delta_{\rm tol}$.

Fig.~\ref{FIG:figure6} presents the transfer-learning of $\eta$ and $N_1$ from channel \#1 ($\tilde{R}=10,~\tilde{W}=1$) to channel \#2 ($\tilde{R}=1,~\tilde{W}=5$). Using the same method for extrapolation in Fig.~\ref{FIG:figure5} with $\beta_\eta = 6.41$ and $\beta_{N_1}=14.19$, the next sampling point determined by Eq.~\eqref{eq:beta} for extrapolation is $\dot\gamma_{p_5}=4.97~{\rm s^{-1}}$, as shown in Fig.~\ref{FIG:figure6}(a2). Subsequently, we perform a DPD simulation of a simple shear flow at $\dot\gamma_{p_5}=4.97~{\rm s^{-1}}$ and have a GPR model for $\eta$ and $N_1$ valid up to $\dot\gamma_{\rm gpr} = 5.63~{\rm s^{-1}}$ as shown in Fig.~\ref{FIG:figure6}(b2), which is sufficient to close the macroscopic SEM solver for channel \#2 whose maximum strain rate is $\dot\gamma_{\max} = 5.08~{\rm s^{-1}}$.
After the flow reaches its steady state, the maximum flow velocity in the classical FENE-P model is $U = 7.97~cm/s$ ($3.74$ in reduced unit) resulting in $Re=3.50$ and $We = 9.31$, while the flow velocity in the macroscale-microscale coupled model is $U = 2.24~cm/s$ ($1.05$ in reduced unit) resulting in $Re = 0.98$ and $We=2.62$.

\subsection{Flow Past a Cylinder}\label{sec:3_2}
The active learning method is now applied to pressure-driven flow past a circular cylinder within a straight channel similar to the previous cases, the channel flow $\#1$ and $\#2$, with walls located at $y=\pm 2L_0$. The inlet is at $x=-10L_0$, the outlet is at $x=15L_0$. A circular cylinder with diameter $D = 2L_0$ placed at the origin, as shown in Fig.~\ref{FIG:figure1}.
The flow is driven by a uniform pressure gradient, $f={dp}/{dx}$, along the channel.
Boundary conditions are imposed on the walls as: the no-slip for velocity, and the Neumann condition for the extra-stress tensor. Periodic conditions are imposed in the streamwise direction.
The cylinder Reynolds number is $Re=UD/\nu_0=2\tilde{R}U/(2.14\tilde{U})$ where the diameter $D=2L_0$, $U$ and $\nu_0=2.14\tilde{\nu}$ are the mean flow velocity and the zero-shear-rate viscosity, respectively. The Weissenberg number is defined as $We=\lambda_0U/D = 0.5\tilde{W}U/\tilde{U}$ where $\lambda_0$ is the relaxation time. Flow past the cylinder is initiated by direct transfer of the constitutive equation as previously learned in the channel flow $\#2$.

\begin{figure}[t!]
\centering
\includegraphics[width=0.6\textwidth]{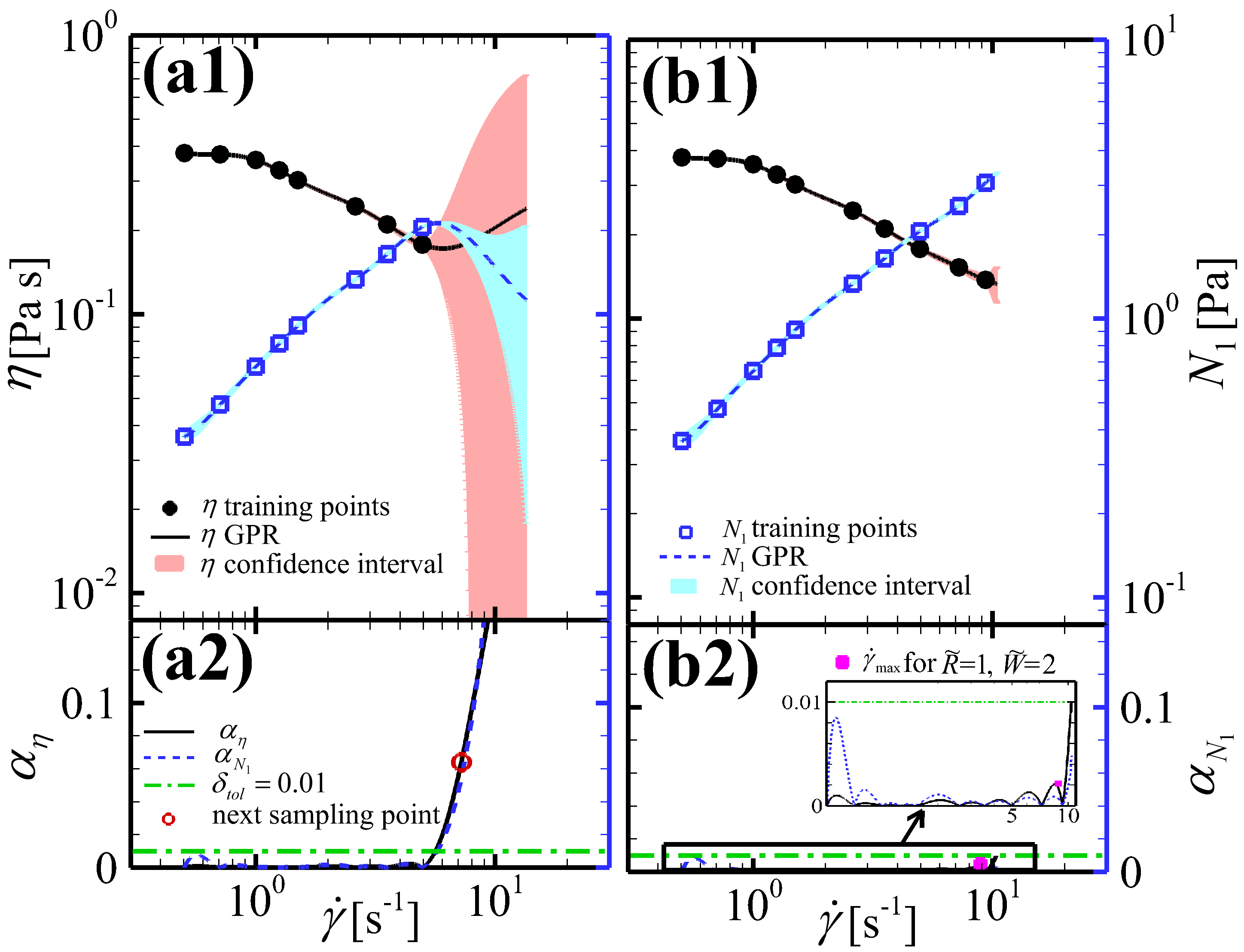}
\caption{Transfer-learning from the channel geometry to flow past a cylinder in a channel geometry. (a) The constitutive closure model learned previously in the channel flow $\#2$, and (b1) the GPR-informed constitutive relations with two more sampling points. In (a1) and (b1) the symbols represent training data while lines are GPR predictions; the shaded area visualizes predicted uncertainties at the $95\%$ confidence interval.
The filled square in (b2) shows the global maximum shear strain rate $\dot\gamma_{\max} = 8.52~s^{-1}$ in the viscoelastic flow at $\tilde{R}=1,~\tilde{W}=2,~\tilde{f}=3$.}
\label{FIG:figure11}
\end{figure}

The constitutive closure shown in Fig.~\ref{FIG:figure6}(b) or Fig.~\ref{FIG:figure11}(a) is valid only up to $\dot{\gamma}_{\rm gpr} = 5.63~s^{-1}$ to satisfy $\max{\alpha}<\delta_{\rm tol}$. As the velocity field around the cylinder develops, the active-learning scheme automatically detects an acceptance violation since the GPR model was constructed for the channel flow $\#2$. This initializes new DPD simulations at extrapolation sampling points $\dot\gamma_{p6}=7.23~s^{-1}$ and $\dot\gamma_{p7}=9.34~s^{-1}$, which are determined by Eq.~\eqref{eq:beta} and $\alpha_\eta(\dot\gamma_{p}) = \beta_\eta\delta_{\rm tol}$ with $\beta_\eta = 6.41$.
Fig.~\ref{FIG:figure11}(b1) shows how the GPR-informed method improves the closure of previous channel flow data $\#2$ by adding sampling points at $\dot{\gamma}_{p6}$ and $\dot{\gamma}_{p7}$, thus yielding an updated constitutive closure valid up to $\dot{\gamma}_{\rm gpr}=10.43s^{-1}$. The filled square in Fig.~\ref{FIG:figure11}(b2) shows the global maximum shear strain rate to be $\dot{\gamma}_{\rm max} = 8.52~s^{-1}<\dot{\gamma}_{\rm gpr}$ for cylinder flow at $\tilde{R}=1,~\tilde{W}=2$ and $\tilde{f}=3$.
This example demonstrates how the learned effective closure constructed in the channel flow can be transferred directly to flow past a cylinder with very few additional expensive DPD simulations needed for a satisfactory closure.

\begin{figure}[t!]
\centering
\includegraphics[width=0.9\textwidth]{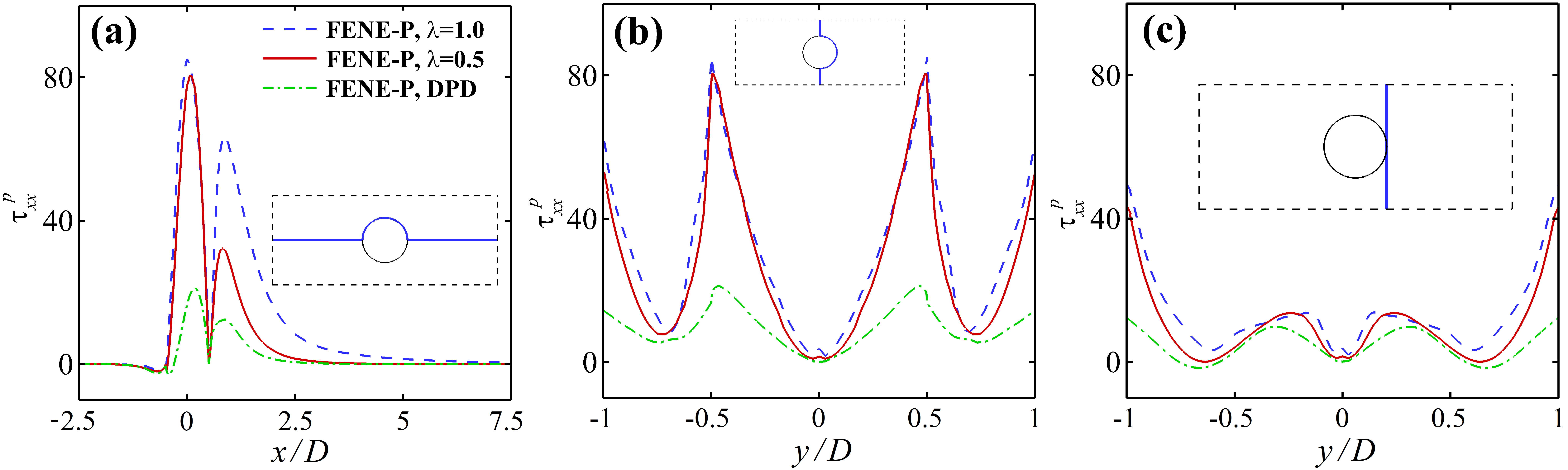}
\caption{Profiles of viscoelastic flow past a circular cylinder. Comparison of normal stress $\tau_{xx}^p$ for two classic FENE-P fluids (FENE-P) and a macroscale-microscale coupled fluid (FENE-P+DPD) along different sections (a), (b) and (c) with the same driving force $\tilde{f}=3$.}
\label{FIG:figure12}
\end{figure}

With a constant uniform driving force $\tilde{f}=3$, the flow past a cylinder in the channel eventually reaches a steady state, where the maximum shear strain rate for the microscale-macroscale coupled system is $\dot\gamma_{\max}=7.12s^{-1}$. The effective viscosity of this shear-thinning fluid is $\nu=1.53\times10^{-4}~m^2/s$ at shear strain rate of $7.12s^{-1}$. With the same driving force $\tilde{f}=3$, we simulate flow past a cylinder confined in a channel for two FENE-P fluids (FENE-P) and a microscale-macroscale coupled fluid (FENE-P+DPD). At their steady states, the maximum flow velocity at the inlet boundary is $U = 3.54~cm/s$ for the classical FENE-P fluid resulting in $Re=2.45$ and $We=2.62$, $U = 1.69~cm/s$ for the microscale-macroscale coupled fluid resulting in $Re=1.17$ and $We=1.25$.
Fig.~\ref{FIG:figure12} compares stress component $\tau_{xx}^p$ for the FENE-P fluids and a macroscale-microscale coupled fluid along different sections of flow past a cylinder under the same driving force $\tilde{f}=3$. The three selected sections for extracting the stress component $\tau_{xx}^p$ are indicated by the blue solid lines in the insets of Fig.~\ref{FIG:figure12}, i.e., section (a) is along the horizontal centerline $\left(y=0\right)$ and on the periphery of the cylinder, section (b) is along the vertical centerline $\left(x=0\right)$ and on the periphery of the cylinder, and section (c) is a vertical line $\left(x=1\right)$ tangent to the cylinder. Results show that the classical FENE-P fluid generates much higher stress $\tau_{xx}^p$ on the solid-wall surfaces than the macroscale-microscale coupled fluid. For a shear-thinning polymer solution, the relaxation time decreases as the shear strain rate increases. The classical FENE-P model that uses a constant relaxation time independent of flow conditions may overestimate the magnitude of stress in the flow past a cylinder. Because the shear strain rate can be large on the no-slip wall surfaces, we observe significant difference of stress on the wall surface and also on the cylinder surface between the classical FENE-P models and the macroscale-microscale coupled model, as shown in Fig.~\ref{FIG:figure12}.

\begin{figure}[t!]
\centering
\includegraphics[width=0.9\textwidth]{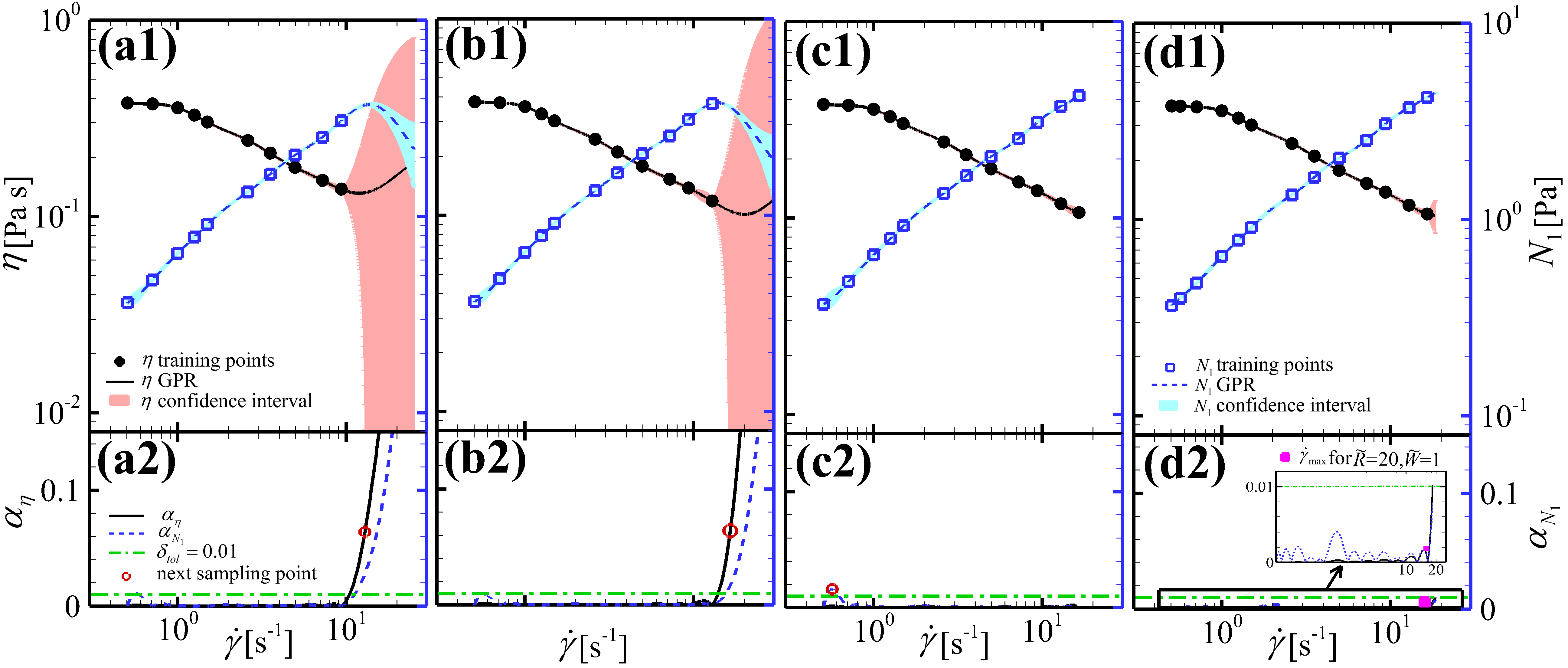}
\caption{Process of constructing a new GPR-informed constitutive closure for $\eta$ and $N_1$ when the flow is changed from $\tilde{R}=1,~\tilde{W}=2,~\tilde{f}=3$ to $\tilde{R}=20,~\tilde{W}=1,~\tilde{f}=0.4$. In (a1-d1), the symbols represent training data while the lines are GPR predictions; the predicted uncertainties are visualized by the $95\%$ confidence interval shown as the shaded area. In (a2-d2) we plot the magnitude of the acquisition functions $\alpha_\eta$ and $\alpha_{N_1}$. The next sampling points $\dot\gamma_{p8}=12.92~s^{-1}$ and $\dot\gamma_{p9}=16.56~s^{-1}$ for extrapolations are determined by $\alpha_\eta(\dot\gamma_{p}) = 6.41\delta_{\rm tol}$, and $\dot\gamma_{p10}=0.57~s^{-1}$ for interpolation. The filled square in (d2) shows the global maximum shear strain rate $\dot\gamma_{\max} = 15.01~s^{-1}$ in the flow at $\tilde{R}=20,~\tilde{W}=1,~\tilde{f}=0.4$.}
\label{FIG:figure13}
\end{figure}

Upon increasing the Reynolds number by changing the flow conditions, the active-learning scheme is able to automatically detect the inaccuracy of the learned constitutive closure, and subsequently it initiates additional DPD simulations for the extra data needed to once again close the microscale-macroscale coupled system. Fig.~\ref{FIG:figure13} shows the process of constructing a new GPR-informed constitutive closure model when the flow is changed from $\tilde{R}=1,~\tilde{W}=2,~\tilde{f}=3$ to another flow past a cylinder at $\tilde{R}=20,~\tilde{W}=1,~\tilde{f}=0.4$ in the simulation, wherein Fig.~\ref{FIG:figure13}(a1) plots the previously learned constitutive relations in the case $\tilde{R}=1,~\tilde{W}=2,~\tilde{f}=3$, and Fig.~\ref{FIG:figure13}(b1) show an updated constitutive relations by adding one sampling point at $\dot\gamma_{p8}=12.92~s^{-1}$ according to Eq.~\eqref{eq:beta}. However, the learned constitutive relations are still insufficient to close the microscale-macroscale coupled system, hence the active-learning scheme adaptively initiates two more DPD simulations at sampling points $\dot\gamma_{p9}=16.56~s^{-1}$ and $\dot\gamma_{p10}=0.57~s^{-1}$ to construct a constitutive closure model valid up to $\dot{\gamma}_{\rm gpr}=18.42s^{-1}$, as shown in Fig.~\ref{FIG:figure13}(d1). Therefore, three more training data are required to once again close the microscale-macroscale coupled system when we change to the flow past a cylinder with $\tilde{R}=20,~\tilde{W}=1,~\tilde{f}=0.4$.

\begin{figure}[t!]
\centering
\includegraphics[width=0.9\textwidth]{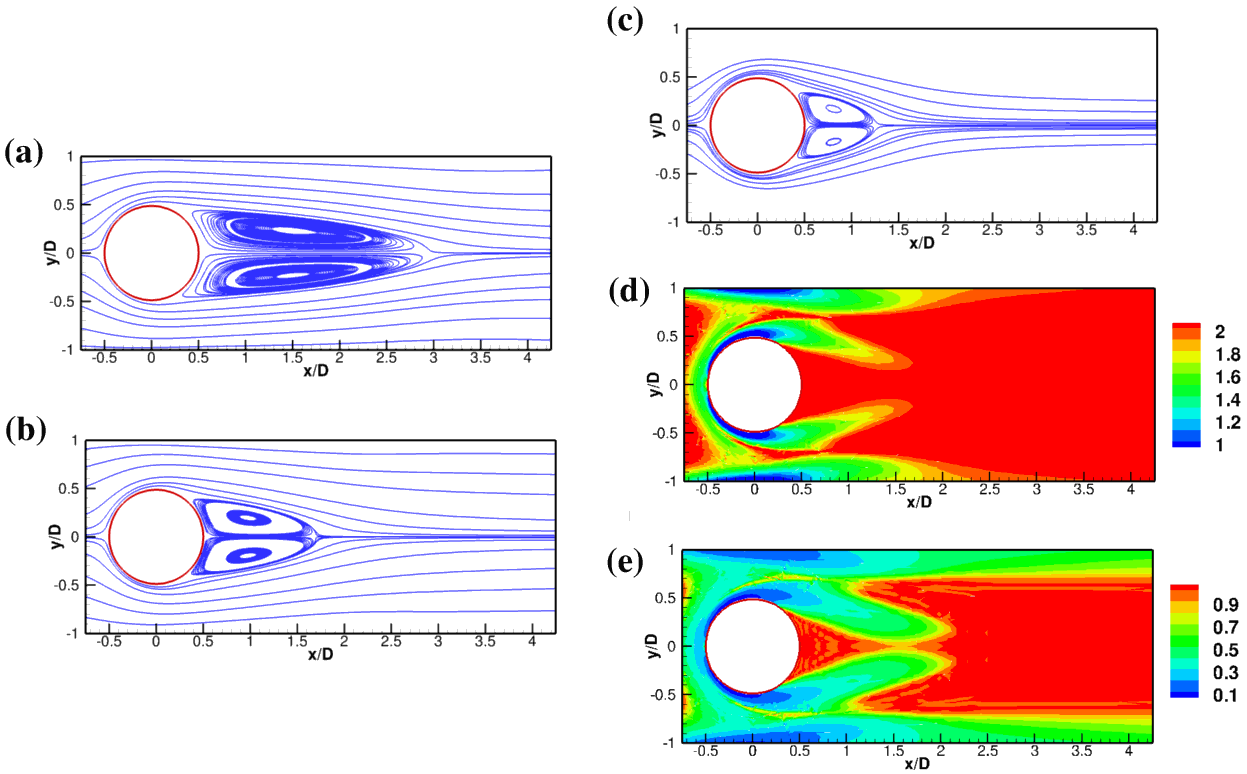}
\caption{Streamlines and contours for $\eta$ and $\lambda$: (a) FENE-P, $\lambda=1.0$ ; (b) FENE-P, $\lambda=0.5$; (c) coupled macroscale-microscale (FENE-P+DPD) at $\tilde{R}=20,~\tilde{W}=1,~\tilde{f}=0.4$; (d) $\eta$ contours normalized by $\eta_0$; (e) $\lambda$ contours normalized by $\lambda_0$.}
\label{FIG:figure14}
\end{figure}

The flow past a cylinder confined in the channel driven by a constant pressure gradient $\tilde{f}=0.4$ with $\tilde{R}=20$ and $\tilde{W}=1$ will eventually reach a steady state, when the maximum shear strain rate for the microscale-macroscale coupled system is $\dot\gamma_{\max}=15.01~s^{-1}$. Considering the shear-thinning effect, the effective viscosity of the fluid is $\nu=1.10\times10^{-4}~m^2/s$ at the shear strain rate of $15.01~s^{-1}$. Using the same driving force $\tilde{f}=0.4$, we carried out simulations of flow past a cylinder in a channel for two classical FENE-P fluids (FENE-P) and a microscale-macroscale coupled fluid (FENE-P+DPD). After the flows reach their steady states, the maximum flow velocity at the inlet boundary is $U = 16.10~cm/s$ for the classical FENE-P fluid resulting in $Re=70.45$ and $We=1.89$, $U = 12.27~cm/s$ for the microscale-macroscale coupled fluid resulting in $Re=53.69$ and $We=1.44$.

Fig.~\ref{FIG:figure14} shows the streamline contours for (a, b) the classical FENE-P viscoelastic model and (c) the coupled macroscale-microscale fluid model for flow past a circular cylinder at $\tilde{R}=20,~\tilde{W}=1$ and $\tilde{f}=0.4$, where a stationary vortex pair behind each cylinder is shown in Figs.~\ref{FIG:figure14}(a-c). The recirculation length $L_w$ is used to quantify the size of a steady separation bubble behind the cylinder. The values of $L_w$ in Figs.~\ref{FIG:figure14}(a),~\ref{FIG:figure14}(b) and~\ref{FIG:figure14}(c) are $2.45D$, $1.20D$ and $0.76D$, respectively, which indicates that under the same driving pressure gradient the wake length in the macroscale-microscale coupled system generates a shorter wake than the classical FENE-P viscoelastic model. This is because the polymer solution possesses a shear-thinning behavior and has larger viscosity in the low shear strain rate regions. The contours of the effective viscosity $\eta$ in the flow past a circular cylinder in steady state are shown in Fig.~\ref{FIG:figure14}(d), and the contours of the relaxation time $\lambda$ are shown in Fig.~\ref{FIG:figure14}(e). We observe that both $\eta$ and $\lambda$ in the macroscale-microscale coupled system are significantly reduced in the vicinity of the cylinder surface and channel walls. Because the friction force on the channel wall surfaces can create an effective confinement suppressing flow separation, we observe that the macroscale-microscale coupled system produces a shorter wake behind the cylinder in~Fig.~\ref{FIG:figure14} compared to the classical FENE-P models.

\begin{figure}[t!]
\centering
\includegraphics[width=0.5\textwidth]{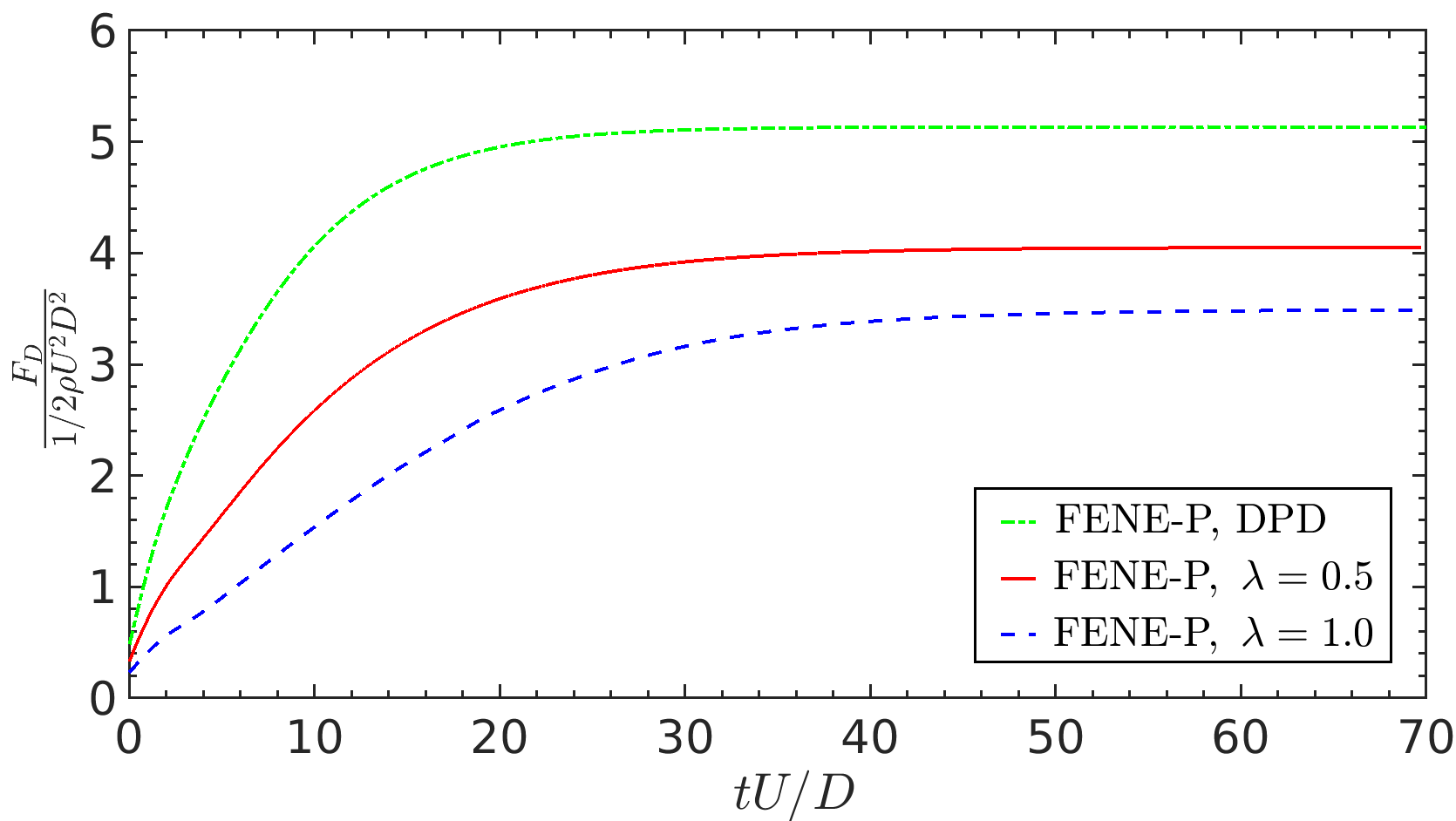}
\caption{Time evolution of the drag coefficient ${F_D}/{(1/2\rho U D^2)}$ at $\tilde{R}=20,~\tilde{W}=1,~\tilde{f}=0.4$ for: (red solid line) FENE-P fluid with $\lambda=0.5$, $U=2.32$, (blue dotted line) FENE-P fluid with $\lambda=1.0$, $U=2.79$, and (green dash-dotted line) the macroscale-microscale coupled fluid (FENE-P+DPD), $U=1.96$. Note that $U$ is the mean velocity calculated at the inlet $x/D=-5$.}
\label{FIG:figure15}
\end{figure}

The drag per unit length on the cylinder is computed from~\cite{1999Dou},
\begin{equation}\label{eq:drag}
F_D=\int_0^{2\pi}\left[ \left(-p+2\eta_s\frac{\partial u}{\partial x}+\tau_{xx}\right) \cos\theta+\left(\eta_s\left(\frac{\partial v}{\partial x}+\frac{\partial u}{\partial y}\right)+\tau_{xy}\right)\sin\theta \right]_{r=R}Rd\theta.
\end{equation}
Fig.~\ref{FIG:figure15} shows the drag coefficient ${F_D}/{(1/2\rho U D^2)}$ for the FENE-P fluids and the microscale-macroscale coupled fluid. The results show that under the same driving force the microscale-macroscale coupled fluid produces larger drag coefficient than the classical FENE-P models, which is consistent with our observations in Fig.~\ref{FIG:figure14} that the classical FENE-P models with constant viscosity and relaxation time produce wakes with longer recirculation length.

\begin{figure}[t!]
\centering
\includegraphics[width=0.9\textwidth]{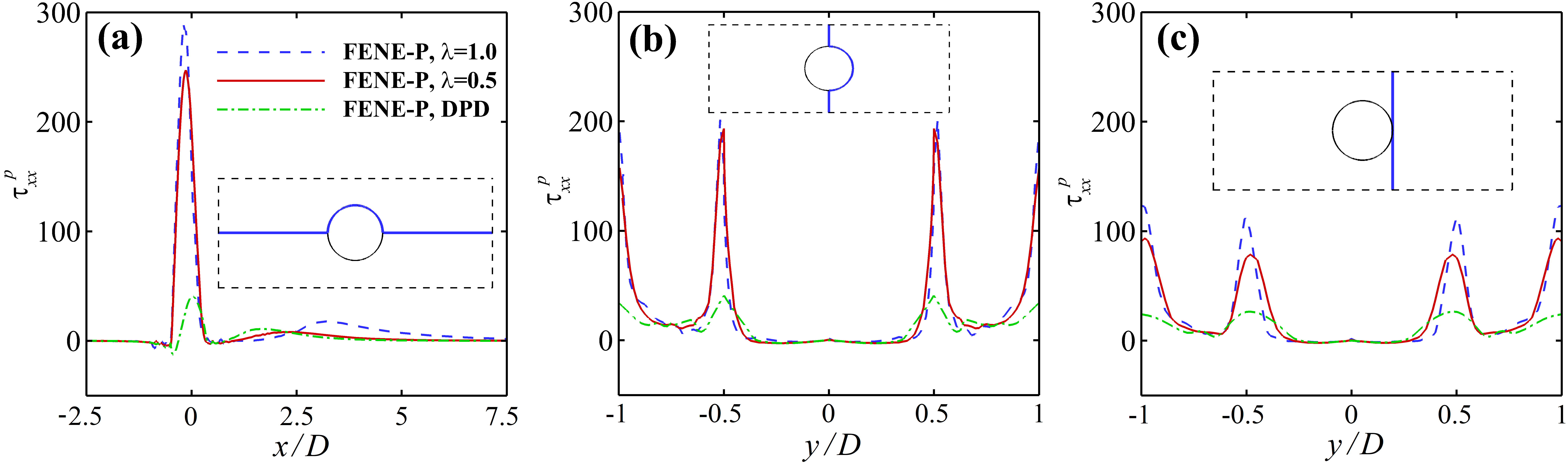}
\caption{Profiles of
normal stress component $\tau_{xx}^p$ for FENE-P for various $\lambda$ and for the macroscale-microscale coupled fluid (FENE-P+DPD) along different sections (a), (b) and (c) at $\tilde{R}=20,~\tilde{W}=1,~\tilde{f}=0.4$.}
\label{FIG:figure16}
\end{figure}

Fig.~\ref{FIG:figure16} compares the normal stress component $\tau_{xx}^p$ for the classical FENE-P fluids and the coupled macroscale-microscale fluid with $\tilde{R}=20,~\tilde{W}=1$ along different sections of flow past a cylinder driven by the same force $\tilde{f}=0.4$. The three selected sections for extracting the stress component $\tau_{xx}^p$ are indicated by the blue solid lines in the insets of Fig.~\ref{FIG:figure16}, i.e., section (a) is along the horizontal centerline $\left(y=0\right)$ and on the periphery of the cylinder, section (b) is along the vertical centerline $\left(x=0\right)$ and on the periphery of the cylinder, and section (c) is a vertical line $\left(x=1\right)$ tangent to the cylinder.
Similar to the previous case of flow past a cylinder, the classical FENE-P fluids generate higher surfaces stress $\tau_{xx}^p$ on the solid-wall than the macroscale-microscale coupled fluid. The decrease of relaxation time with shear-rate of the latter renders it less elastic than the FENE-P model with its constant relaxation time which overestimates stress levels in the flow past a cylinder. Fig.~\ref{FIG:figure16} shows significant differences of normal stress $\tau_{xx}^p$ on both the wall surfaces and on the cylinder surface between the standard FENE-P model and the macroscale-microscale coupled model.

\section{Summary and Discussion}\label{sec:4}
The macroscale simulation of viscoelastic flows requires a constitutive equation for the stress tensor whose parameters are usually determined from physical rheological tests, so that the flow problem solution begins with expensive laboratory work rather than a simple handbook lookup.
This investigation has explored an alternative approach to enable viscoelastic flows to be initiated without physical testing, namely by the use of microscale numerical models.
The macroscopic viscoelastic response to deformation of a polymeric fluid is a manifestation of the microstructural dynamics of polymer molecules, which results in bulk rheology described by complicated nonlinear relationships between the stress tensor and the strain rate. In this work the fluid is taken to be an aqueous polyacrylamide solution for which experimental steady shear data has been used to fit the parameters of the FENE-P constitutive equation.
Hence, the starting point for the active- and transfer-learning scheme guided by Gaussian process regression (GPR) is the selection of a viable microscopic model as the alternative to experimental testing. Here it is taken to be dissipative particle dynamics (DPD) applied to bead-spring chains suspended in a DPD solvent, which generates realistic bulk rheology consistent with the known experimental data.
In the current work, DPD explicitly simulates the chain dynamics of groups of polymer molecules in flows at prescribed shear rates the results of which yield the stress tensor 
computed by the Irving-Kirkwood formula.
Numerical experiments showed that the standard DPD polymer chains of beads with constant cutoff radius to be inadequate to capture the shear-thinning viscosity of the experimental polymer solution.
The DPD model was modified to allow the bead cutoff radius to vary with the second invariant of local strain rate. The model then captures the strain rate dependence of the viscosity and normal stress consistent with the experimental data.

The macroscale equations of continuity and momentum together with the constitutive equation were solved by the spectral element method (SEM).
To enhance the stability of the SEM solver a nonlinear diffusive term  was added the constitutive equation similar to entropy viscosity method in aerodynamic flows, which significantly stabilizes the SEM solution of viscoelastic flows at large Weissenberg ($We$) number.
The macroscale counterpart to the new DPD model with strain-rate dependent cutoff radius is the modified FENE-P model with polymer viscosity and relaxation-time dependent on the second invariant of strain rate analogous to the White-Metzner model with variable viscosity and relaxation time.

In principle, the SEM solver should require the DPD solver to return a stress tensor for each element, and hence we need to perform as many DPD simulations as the number of elements in the SEM solver. However, DPD is a stochastic method, and it is relatively computational expensive to extract a statistically meaningful stress tensor from DPD simulations. We cannot afford carrying out one DPD simulation on each SEM element, and we need a regression model to predict the effective viscosity and relaxation time based on an affordable number of DPD simulations. To minimize the number of necessary DPD runs, we introduced an active-learning strategy to couple the DPD solver and the SEM solver. More specifically, in the microscale-macroscale coupling, the SEM solver provides local transient flow field to initiate DPD simulations of polymer solution, while the DPD solver returns the stress tensor at selected shear strain rates, wherein an active-learning scheme built on Gaussian process regression was used to learn the dependence of viscosity and relaxation time on the shear strain rate. By defining a proper acquisition function and an acceptance criterion, the training points can be adaptively selected so that the DPD simulation is performed only when it is really necessary.

Flow in a straight channel between two parallel plates was used as an example to demonstrate the active-learning procedure in the microscale-macroscale coupled simulations of viscoelastic flows. In this case, informed by the GPR,
the optimal sampling points were adaptively selected on-the-fly to show that the active-learning scheme applied to the multiscale simulation of channel flow required only seven DPD simulations to build a valid constitutive closure model, while the range of strain rate was updated as the flow field evolved in time. For viscoelastic flows using the same working fluid, the learned constitutive closure model for one specific viscoelastic flow can be directly transferred to other flow problems in different domains or to the same problem at different flow conditions. In the case of a viscoelastic fluid flowing through a channel, we demonstrated  how the GPR model constructed in a channel flow can transfer to another channel flow with different flow velocity and viscoelasticity. Also, we demonstrated that the GPR model constructed in channel flows can directly be transferred to the viscoelastic flow past a circular cylinder by actively initializing a few more DPD simulations.

Active- and transfer-learning for multiscale modeling is a new paradigm, which is readily applicable to other microscale-macroscale coupled simulations of complex fluid phenomena. Furthermore, it can be seamlessly implemented using our open source multiscale universal interface (MUI)~\cite{2015Tang}.
In the present study, we designed the DPD system in steady shear flows, and only extracted the shear dependent quantities, such as shear-thinning viscosity and the normal stress difference. It would be very interesting to study other microstructure-dependent viscoelastic properties, such as the elongational viscosity and the viscoelastic response in extensional flows~\cite{1995Bird}. In microscale particle-based simulations, simple shear flows can be easily generated using moving plates or the Lees-Edwards boundary condition~\cite{2016Pan}. However, it is still unclear how one can create a steady elongational flow in particle-based simulations and extract extension-related rheology properties.
Moreover, we only considered the rheology data obtained from DPD simulations to construct the effective constitutive closure model to close the continuum momentum equation. When there are different sources of rheology data from either different computational models or different experiments, the idea of applying multi-fidelity for multiscale modeling would be effective, where the complicated cross-correlations between different data with various fidelities can be found by training a deep neural network~\cite{2019Meng,2019Tipireddy}.

\section*{Acknowledgements}
This work was also supported by DOE PhILMs project (No.\ DE-SC0019453)
and by the Army Research Laboratory (W911NF-12-2-0023).
This research was conducted using computational resources and services at the Center for Computation and Visualization, Brown University.

\appendix
\section{Entropy-viscosity method }\label{sec:app}
In order to stabilize the SEM simulation at high Weissenberg numbers, the entropy viscosity method (EVM) is employed. EVM introduces a nonlinear diffusion term to the FENE-P constitutive equation, given by:
\begin{equation}
    \overset{\triangledown}{\mathbf{C}}=-\frac{1}{\lambda}\frac{L^2}{L^2-3}\left(g\mathbf{C}-\mathbf{I}\right)+\nabla\cdot\left(\nu_t\left(\nabla\mathbf{C}+\nabla^T\mathbf{C}\right)\right),
\end{equation}
in which $\nu_t$ is the so called entropy viscosity, and it can be computed in each element $K$ at the collocation points $ij$:
 \begin{equation}\label{eq:nut}
\nu_t|_K=\min\left\{ \alpha_1 \|u\|_{L^\infty\left(K\right)} \delta_K,\alpha_2\frac{\|R_{ij}^K\left(\mathbf{C}\right)\|_{L^\infty\left(K\right)}}{\|E_{ij}^K\left(\mathbf{C}\right)-\overline{E}\left(\mathbf{C}\right)\|_{L^\infty\left(\Omega\right)}}\delta_K^2\right\},
\end{equation}

\begin{equation}
   E_{ij}^K\left(\mathbf{C}\right)=\frac{1}{2}\left(\mathbf{C}_{ij}^K-\overline{\mathbf{C}}\left(\Omega\right)\right)^2,
\end{equation}

\begin{equation}
    \overline{E}\left(\mathbf{C}\right)=\frac{\int_\Omega E_{ij}^K\left(\mathbf{C}\right) \cdot \,dX}{\int_\Omega\,dX},
\end{equation}

\begin{equation}
  R_{ij}^K\left(\mathbf{C}\right)=\mathbf{C}\cdot\left(\dot{\mathbf{C}}+\frac{1}{\lambda}\frac{L^2}{L^2-3}\left(g\mathbf{C}-\mathbf{I}\right)\right)\Bigg|_{ij}^{K},
\end{equation}
where $\delta_K$ is the minimum distance between two quadrature points in element $K$, $\overline{\mathbf{C}}\left(\Omega\right)$ are the mean values of $\mathbf{C}$ in the whole region $\Omega$, $\alpha_1$ and $\alpha_2$ are two parameters.

\begin{figure}[t!]
\centering
\includegraphics[width=0.45\textwidth]{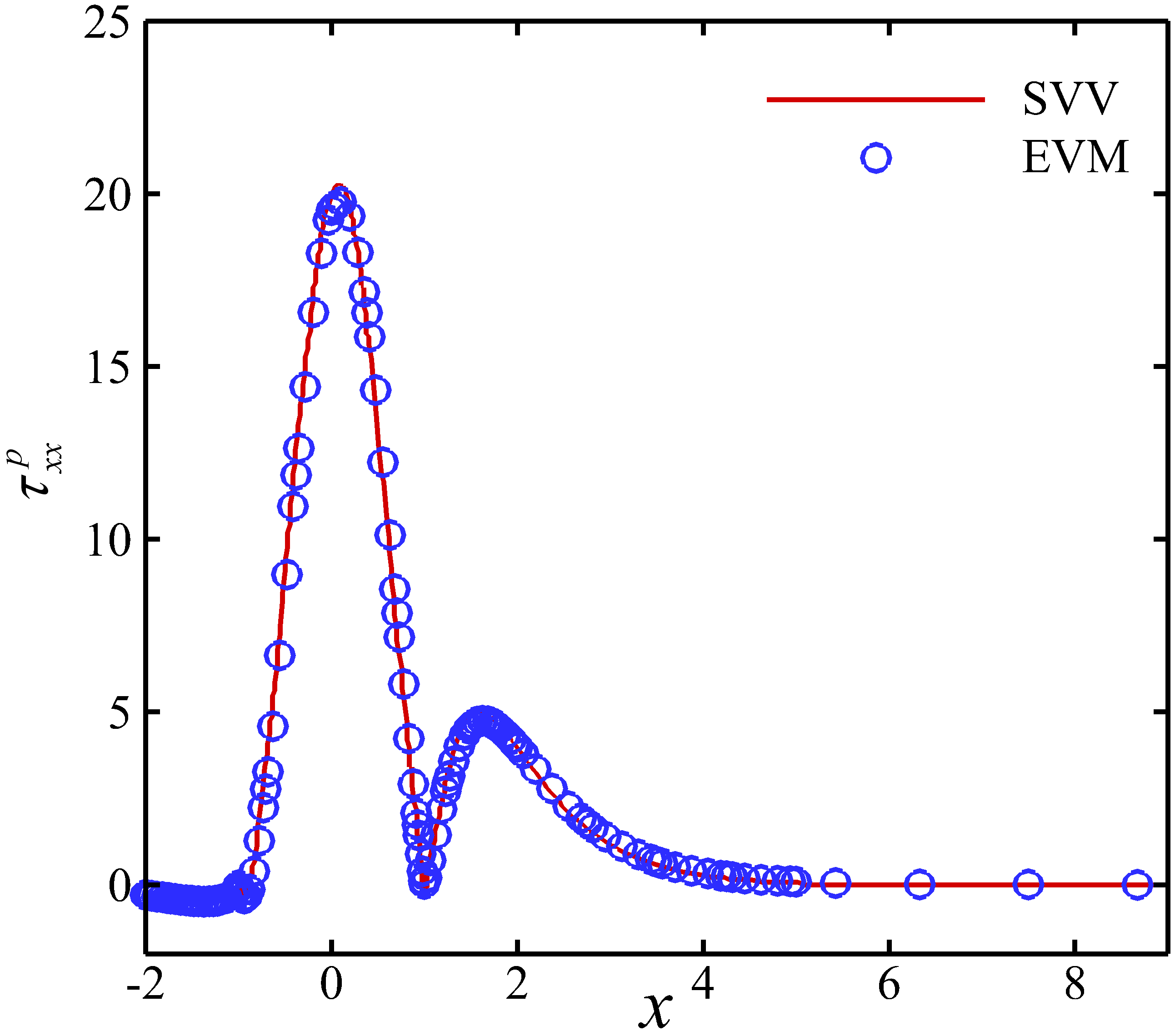}
\caption{Verification of EVM against the SVV method in a flow past a circular cylinder for polymer stress $\tau^p_{xx}$.}
\label{FIG:figure10}
\end{figure}

A benchmark test for the viscoelastic flow past a cylinder was chosen to validate the EVM method with the same geometry given in ~\cite{2003Ma}, where the channel dimensions are $\left[-20L_0,20L_0\right]\times\left[-2L_0,2L_0\right]$, and the cylinder with radius of $L_0$ is placed at $\left(x_0,y_0\right)=\left(0,0\right)$. Fig.~\ref{FIG:figure10} shows the verification of the proposed EVM against the spectral vanishing viscosity~(SVV) method~\cite{2003Ma} for the polymer axial normal stress $\tau^p_{xx}$ along the axis of symmetry $\left(y=0\right)$ and on the periphery of the cylinder. The amplitude $\varepsilon$ and cut-off wave number $M$ in the SVV simulation are $\varepsilon=0.001$ and $M=2$. The parameters $\alpha_1$ and $\alpha_2$ in Eq.~\eqref{eq:nut} for the EVM simulation are $\alpha_1=0.001$ and $\alpha_2=0.1$. These results verified that the proposed EVM method can generate a stress profile $\tau_{xx}^p$ as good as that of the SVV method but EVM produces more stable solutions.

\end{document}